\documentclass[11pt,oneandhalf,chaparabic]{metu}

\usepackage{appendix}
\usepackage{graphicx}
\usepackage{algorithm}
\usepackage{algorithmic}
\usepackage{amsmath}
\usepackage{txfonts}
\usepackage{subfigure}
\usepackage{amssymb}
\usepackage{subfigure}
\usepackage{multirow}
\usepackage{colortbl}
\usepackage{slashbox}

\usepackage{hhline}
\newtheorem{theorem}{Theorem}[section]
\newtheorem{definition}[theorem]{Definition}

\newenvironment{defn*}{\begin{definition}}{\end{definition}}

\author{Kamil \c{C}{\i}nar}
\authoruppercase{KAM\.{I}L \c{C}INAR}
\graduateschool{Natural and Applied Sciences}
\shortdegree{M.Sc.}
\director{Prof. Dr. Canan \"Ozgen}
\headofdept{Prof. Dr. Sinan Bilikmen}

\title{DESIGN AND CONSTRUCTION OF A MICROWAVE PLASMA ION SOURCE}
\degree{Master of Science}
\department{Physics}
\date{February 2011}
\supervisor{Prof. Dr. Sinan Bilikmen}
\departmentofsupervisor{Physics Dept., METU}
\keywords{Ion Source, Microwave Plasma, Dense Plasma, Ion Generation, Ion Extraction}

\turkishtitle{ M\.{I}KRODALGA PLAZMA \.{I}YON KAYNA\u{G}I TASARIMI VE \"URET\.{I}M\.{I}}
\turkishdegree{Y\"uksek Lisans}
\turkishdepartment{Fizik B\"ol\"um\"u}
\turkishdate{\c{S}ubat 2011}
\turkishsupervisor{Prof. Dr. Sinan Bilikmen}
\anahtarklm{\.{I}yon Kaynaklar{\i},  Mikrodalga Plazma, Yo\u{g}un Plazma, \.{I}yon \"Uretimi, \.{I}yon \c{C}{\i}kart{\i}m{\i}}

\committeememberii{Prof. Dr. Sinan Bilikmen}
\committeememberi{Assoc. Prof. Dr. Serhat \c{C}ak{\i}r}
\committeememberiii{Assoc. Prof. Dr. \.{I}smail Rafatov}
\committeememberiv{Dr. Burak Yedierler}
\committeememberv{Dr. Ali Ala\c{c}ak{\i}r}
\affiliationi{Physics Department, METU}
\affiliationii{Physics Department, METU}
\affiliationiii{Physics Department, METU}
\affiliationiv{Physics Department, METU}
\affiliationv{Research and Development Department, SANAEM}

\begin{document}

\begin{preliminaries}
  \maketitle
  \makeapproval
  \plagiarism
  \setlength{\parindent}{0em}
  \setlength{\parskip}{10pt}
% Abstract
  \begin{abstract}\oneandhalfspacing
	 This thesis is about the  designing and constructing a microwave ion source. The ions are generated in a thermal and dense hydrogen plasma by microwave induction. The plasma is generated by using a microwave source with a frequency of 2.45 GHz and a power of 700 W. The generated microwave is  pulsing with a frequency of 50 Hz. The designed and constructed microwave system   generates  hydrogen  plasma in a pyrex plasma chamber. Moreover, an ion extraction unit is designed and constructed in order to  extract the ions from the generated hydrogen plasma. The ion beam extraction is achieved and ion currents are measured. The plasma parameters are determined by a double Langmuir probe and the  ion current is measured by a Faraday cup. The designed ion extraction unit is simulated by using the dimensions of the designed and constructed ion extraction unit in order to trace out the trajectories of the extracted ions. 

  \end{abstract}
% Oz
  \begin{oz}\oneandhalfspacing
	Bu tezin konusu mikrodalga iyon kayna\u{g}{\i} tasar{\i}m{\i} ve \"uretimidir.  \.{I}yonlar, yo\u{g}un ve termal hidrojen plazmas{\i} i\c{c}inde \"uretilmi\c{s}tir. Plazma, 2.45 GHz frekansl{\i} ve 700 Watt \c{c}{\i}k{\i}\c{s} g\"u\c{c}\"unde mikrodalga kayna\u{g}{\i} kullan{\i}larak olu\c{s}turulmu\c{s}tur. \"Uretilmi\c{s} mikrodalga kayna\u{g}{\i}, 50 Hz'lik  frekans at{\i}ml{\i} mikrodalga  olu\c{s}turmaktad{\i}r. Tasarlanan ve \"uretilen mikrodalga sistemi, payreks plazma kab{\i} i\c{c}inde hidrojen plazmas{\i} \"uretmi\c{s}tir. Ayr{\i}ca iyonlar{\i} olu\c{s}turulan hidrojen plazmas{\i}ndan \c{c}{\i}karmak i\c{c}in    iyon \c{c}{\i}karma \"unitesi tasarlan{\i}p \"uretilmi\c{s}tir. \.{I}yon   \c{c}{\i}karma ba\c{s}ar{\i}lm{\i}\c{s}t{\i}r.       Plazma parametreleri   \c{c}ift Langmuir sondas{\i} ile, iyon ak{\i}m{\i} ise Faraday kab{\i} ile \"ol\c{c}\"ulm\"u\c{s}t\"ur. Tasarlanan iyon \c{c}{\i}karma \"unitesi boyutlar{\i} kullan{\i}larak iyon y\"or\"ungeleri sim\"ule edilmi\c{s}tir.

  \end{oz}
% Dedication
  \dedication{\textit{To my parents}}
  \setlength{\parindent}{0em}
  \setlength{\parskip}{10pt}
% Acknowledgment
  \begin{acknowledgments}\oneandhalfspacing
I express my sincere appreciation to Prof. Dr. Sinan Bilikmen who provided the opportunity of working on the present subject. I respectfully acknowledge Dr. Ali Ala\c{c}ak{\i}r for this continuous support on not only proffessional issues but also on personal matters. His unique effort on creating  a  very productive working environment  will always be appreciated. Moreover, my special thanks goes to Dr. Erdal Recepo\u{g}ulu and Dr. Hande Karadeniz for their invaluable support.
%I would like to thank to my supervisor Prof. Dr. Sinan Bilikmen for providing me opportunity to work on this subject. And I would like to show my repects and thank to Dr. Ali Ala\c{c}ak{\i}r for his physical and psychological support while working on this experimental study and  for creating a very productive working environment. Moreover, I would like to thank to Dr. Erdal Recepo\u{g}ulu and Dr. Hande Karadeniz for their great help.
  \end{acknowledgments}
  \setlength{\parindent}{0em}
  \setlength{\parskip}{3pt}
% Table of Contents (do not try to edit ) 
  \tableofcontents

\listoftables

\listoffigures

\end{preliminaries}

\setlength{\parindent}{0em}
\setlength{\parskip}{10pt}

% CHAPTER 1
\chapter{INTRODUCTION}
Ion sources started to emerge while Goldstein was working on the canal rays in 1886 before  designing low-current ion sources, which  electron-atom collision mechanisms were used for \cite{wolf}. In 1930's, by investigating the arc discharge, higher ion currents started to be provided.  RF and microwave discharges started to be investigated  during the next decade and  they were used in production of ion beams \cite{wolf}. As it is seen that the origins of the ion sources are the atomic-nuclear physics research and ion implantation for microelectronic applications; moreover,  ionization sources were developed for space propulsion applications  in  1960's \cite{roth}. After all, ion sources have become indispensable parts of particle accelerators and ion implantation systems.
For areas of usage, there are many types of ion sources with different working mechanisms; on the contrary, electron sources have limited variety \cite{wolf, roth}. The ion types can be determined for the  corresponding applications. Ion sources are mainly used to produce mono-energetic and unidirectional ion beams \cite{roth}. The generated beams are utilized by guiding them to  ion beam lines or by direct guiding them to application zones.

There are plenty types of ion sources. The variety of the ion sources arises from the different ways of  ion generation from solids, liquids and gases and also the variety of generating plasma such as DC discharge, arc discharge, RF discharge, microwave discharge and laser driven plasmas \cite{roth}. The general types of ion sources  are listed  in Table \ref{typical ion sources} and these ion sources can be divided into groups with respect to the ways of ion generation and areas of applications \cite{wolf}.

\begin{table}[H]
\caption{Typical Ion Sources}
\begin{center}
\begin{tabular}{|l|}
   \hline
  Cathodes \\
  Electron Bombardment Ion Sources \\
  Plasmatron Ion Sources \\
  Magnetron and Freeman Types Ion Sources \\
  Penning Ion Sources \\
  Multicusp Ion Sources \\
  RF Ion Sources \\
  Microwave Ion Sources \\
  ECR Ion Sources \\
  Laser Ion Sources  \\
  Electron Beam Ion Sources /Trap \\
  Vacuum Arc Ion Sources \\
  Large Area Ion Sources\\
  Industrial Ion Sources\\
  Liquid-Metal Ion Sources\\
  Polarized Ion Sources\\
  Cluster Ion Sources\\
  Ion Diodes\\
  Ion Sources for Space Applications\\
  \hline
\end{tabular}
\end{center}

\label{typical ion sources}
\end{table}

The ion sources or ion beam generators have a wide variety of units, such as  plasma induction systems,  ion extraction systems, electronic units that  induce  plasma and supply the extraction voltage. There are also units for guiding the ion beams.  The ion sources necessitate an interacting multi-discipline study which should have the knowledge of plasma physics, electrical and electronics engineering, and computational systems. 

In general, at the end of the ion beam line, there is a linear accelerator. The produced ion sources are connected to the open-end of this accelerator via vacuum cones. The accelerated ions are focused by an Einzel lens and dispersed by  two quadrupole magnets. Moreover, a dipole magnet is used in order to deflect the accelerated ions, passing through the ion beam line. At the other end of the ion beam line, there is a target which is subjected to ion bombardment.

In this thesis, microwave ion sources have been investigated for the  design and the construction. The designed and produced ion sources have been used for the ion beam lines.

In 1977, the first ion source was built by Sakudo and he also   developed  a microwave ion source with slit extraction \cite{wolf}. Ishika, in 1984, used permanent magnets in order to built a compact microwave ion source \cite{wolf}. Afterwards the gradually increasing interest on  microwave ion sources has spread over, because the microwave ion sources can reach high current densities at low pressure plasmas. 

The built-in microwave ion source system does not contain any magnetic confinement or any magnetic support for the plasma, like the electron cyclotron resonance (ECR) microwave discharges.

The organization of the thesis is as follows:
In chapter 2, the theory of ion production via  plasma is discussed and the mechanisms of the direct current (DC) plasma and the microwave (MW) plasma are reviewed. In chapter 3, the mechanism of the ion extraction is  considered. In addition,  the plasma (space) potential theory and the vacuum gradient are discussed. Characterization of ion beams and ion sources are explained in Chapter 4.  The main part, the designed and constructed microwave plasma ion source is  expressed in Chapter 5 where  the produced parts of the  system  are explained. As the last chapter, Chapter 6, the data analysis of the ion source is done and the theoretical values of the designed system are compared with the simulation of the designed system and the measured data. Finally results and outcomes of the thesis are concluded.   

\newpage

% CHAPTER 2
\chapter{ION PRODUCTION VIA PLASMA}
\label{chp:ion production via plasma}
Although there are many ways of producing ions, only ion production mechanism via plasma is the concern of this chapter.  Ion production process can be divided into two parts. The  first one is the production of raw ions from a generated plasma  and the second  is the extraction of the raw ions from the generated plasma. 

\section{PLASMA}
%\section{PLASMA GENERATION MECHANISMS}
 In the universe, more than 99\% of observed matter comprises plasma such as interstellar matter, nebulea, supernova, stars and also the flame \cite{konuma}. Plasma is the fourth state of  matter. If gases are heated,  electrons of the gas molecules start to oscillate and the probability of ionization of the gas increases.  In addition,  electrons  acquire more energy  and their coupled atoms  break up and move around freely. Consequently, the electrons prevail the electrostatic forces.  If  a quasineutral gas  consists of charged and neutral particles which exhibits collective behaviour, this quasineutral gas is named   plasma \cite {chen}. 

 Quasineutrality is defined as approximately equal numbers of negative and positive charges existing in a system. 

 The collective behaviour of the plasma can be described by firstly comparing with the kinetic theory of gases. The kinetic theory of  gases is that no net forces; such as, electromagnetic forces which act upon the  gas particles. Because the gas particles  carry no net charges, they are neutral. So the particles move in straight lines before they collide each other with  a distribution of velocities (Gravitational forces are not our concern at this moment.) \cite{grill}. In plasma, charged particles create long range electromagnetic forces. Dominating  motions of the charged particles and local charge concentrations affect the whole plasma even  if the local charge concentration is far from the concerning area. While the neutral particles exhibit straight  line motions after their collisions with each other, the charged particles exhibit continuously changing motions because of electromagnetic forces between each other \cite{grill}. The electromagnetic forces are long-range forces and  interactions which occur between the charged particles  last. These continuous interactions of the charged particles do exhibit nonlinear behaviours. If there is a  locally varied charges or a varied current distribution, there will be impacts on the whole plasma; i.e., the affected charged distribution will evolve into a new charged distribution.

 In addition to the quasineutrality and collective behaviour of the plasma,  the temperature, the number density of the particles and the most importantly, Debye length should be considered while defining the plasma. 

 Before defining the Debye length, the Debye shielding should be defined. The Debye shielding is the ability of the plasma to shield out electric potentials that are applied to the plasma \cite{chen}. The electric  potentials can be produced by local charges or by inserting electrodes inside the plasma. The distance, where the potential vanishes, is called the Debye length. The Debye length, $\lambda_D$, is given by the expressions below \cite{wolf}.

\begin{equation}
 \lambda_{D}^{2}=\dfrac{{\varepsilon_{0}} kT_{e}}{e^{2}n_{e}} \quad,
\end{equation}
\begin{equation}
 \lambda_{D}=743\sqrt{\frac{T_{e}}{n_{e}}} \quad,
\end{equation}

where $k$ is the Boltzmann constant,  $\varepsilon_{0}$ is the vacuum permittivity and $e$ is the elementary charge ($\sim 1.6\times10^{-19}$C (Coulombs)). The parameters, $T_{e}$ and $n_{e}$ are the electron temperature in electron volts,  the electron number density in inverse cubic centimeter,  respectively and the Debye length, $\lambda_{D}$ is in centimeters \cite{wolf}. If the Debye length $\lambda_{D}$ increases, the density of the plasma will decrease; moreover, the Debye length, $\lambda_{D}$, increases with  increasing $kT_{e}$. The unit of $kT_{e}$ is Joule.   

The Debye length also characterizes the plasma with additional parameters such as, $N_{D}$, which is the number of particles in the the Debye region \cite{chen}.  $N_{D}$ is computed by  Equation 2.3 for a Debye sphere which a sphere with a radius of $\lambda_{D}$.
\begin{equation}
 N_{D}=\frac{4}{3}n_{e}\pi\lambda_{D}^{3}
\end{equation}

 The gas is considered as a plasma when the three conditions, below, are satisfied \cite{chen}.
\begin{equation}
 \lambda_{D}<<L \quad,
\end{equation}
where $L$ is the size of the plasma system. The dimensions of the plasma region should be much larger than the Debye length.
  \begin{equation}
 N_{D}>>1 
\end{equation}
The number of particles should be much more than one in order to consider bulk of the charged  particles as plasma.  
\begin{equation}
 \omega_{p}\tau>1
\end{equation}

 Equation 2.5 is the requirement of the collective behavior.  In Equation 2.6, $\omega_{p}$ is the frequency of the plasma oscillations and $\tau$ is the mean time between the collisions with neutral atoms. The plasma frequency should be  more than the mean time of the collisions with the neutrals.

 There are various ways  to generate a plasma; however, two ways of generating plasma; such as DC (Direct Current) discharge,   and MW (Microwave) discharge, have been mentioned respectively in this thesis. However, the microwave discharge have been discussed in detail.

\section{DC DISCHARGE}
% \subsection{DC DISCHARGE}

\begin{figure}[H]
 \centering
 \includegraphics[width=0.5\textwidth]{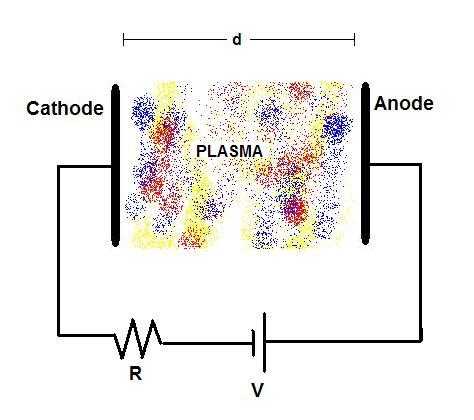}
 \caption{Schematic of a DC discharge}
\label{Schematic of a DC discharge}
\end{figure}

 A DC  discharge can be achieved by applying a DC voltage between two conducting electrodes which are inserted into a gas at low pressures \cite{grill}. The typical gas pressure is in between    0.1 torr (0.133322 mbar) and 10 torr (13.3322 mbar) in order to process DC discharge plasma \cite{konuma}.  A schematic drawing of the set-up is given in Figure \ref{Schematic of a DC discharge}.

Any gas medium contains free ions and free electrons which  arise from  interactions between cosmic rays, environmental radiation with the gas atoms or a result of field emission from  any violence  on the surface, where electric fields are strong \cite{konuma,grill}. These  free charge carriers are accelerated through these electric field lines and they start to collide with the atoms or molecules, encountered with these free charge carriers \cite{konuma}.

These small amount of accelerated free charge carriers loses their kinetic energy by  collisions; however, the voltage, applied to the electrodes, maintain these particles to build up their kinetic energies in order to ionize or excite targets that are the neutral particles, such as atoms and molecules \cite{grill}. If the kinetic energies of the free charge carriers are enough to excite or ionize the neutral particles, the accelerated charge carriers lose their kinetic energies by making inelastic collisions with the neutral atoms or molecules; otherwise, the collisions are elastic if the energy of the free charge carriers have too low  to excite or ionize the targets which are the atoms and the molecules \cite{grill}. (Ionization is to break off electrons from the atoms whereas excitation is to bring electrons of an atom to a higher energy level than the electrons have.)  With the ionization and excitation of the atoms or molecules, the number of produced charge carriers increases and an avalanche effect occurs. Besides of the free charge carriers, the produced ions and electrons contribute the excitation or ionization process in the medium. The produced ions and especially the produced electrons gain kinetic energy by the applied voltage. If there are elastic collisions between electrons and targets, the loss of the electron energy or the transferred energy  can be calculated by Equation 2.7 \cite{grill}:

\begin{equation}
 W_{tr}=\dfrac{2m_{e}}{M_{T}}W_{e} \quad,
\end{equation}

where $W_{tr}$ is the transferred energy  to a target by an electron, $M_{T}$ is the mass of the target, $m_{e}$  is the mass of an electron and $W_{e}$ is the energy of the electron.
For the inelastic collisions between the energetic electrons and the heavy target, the mechanism and the calculation of the collision change. The average fraction of the transferred energy is given by Equation 2.8 \cite{grill}:

\begin{equation}
 \dfrac{W_{tr}}{W_{e}}=\dfrac{M_{T}}{m_{e}+M_{T}}
\end{equation}

 This avalanche effect is caused by the new produced electrons and ions. An electron multiplication process takes place. The number of electrons $N$, flowing through the anode electrode in a unit time, can be calculated by Equation 2.9 \cite{konuma} 

\begin{equation}
 N=N_{0}\frac{e^{\alpha d}}{1-\gamma(e^{\alpha d}-1)} \quad,
\end{equation}

 where $N_{0}$ is the initial number of electrons, $\alpha$ is the probability that the accelerated electron ionizes a gas atom as it travels   a unit distance in the discharge tube ($\alpha$ is  the first Townsend coefficient) \cite{konuma,tarey}. The term $e^{\alpha d}$ is the amplification factor at a distance of $d$. Beside of this, it is also the number of ions, produced by the primarily electrons. The coefficient $\gamma$ is the efficiency of secondary electrons,e.i., the secondary electrons are produced by the ions, striking the cathode electrode. It is also called as the second Townsend coefficient \cite{konuma,tarey}. The secondary electron emission coefficient  $\gamma$ is determined by the type of material, which  the cathode is made of, and the structure of the cathode surface; in addition to them, type of gas and reduced electric field $E/p$ have  importance \cite{fridman}. Here, $p$ is the pressure of the region where the plasma is formed.

 The rapid transition, which is from a very poor electrical conductor with the resistivity of $\sim 10^{14} \Omega m$ to a relatively good conductor with a resistivity that is many orders of magnitude lower, characterizes the breakdown mechanism for the concerning gas, which is  in a  tube \cite{konuma}.  When the value of $N$ goes to infinity, the electrical breakdown of the gas  occurs  in the gap between the two electrodes \cite{konuma}. The electron amplification factor $\frac{N}{N_{0}}$, goes to infinity if the  denominator goes to zero in Equation 2.9.  The condition of the electrical breakdown is given by Equation 2.10  \cite{konuma}.

\begin{equation}
 \gamma (e^{\alpha d}-1)=1
\end{equation}

 In order to sustain the DC plasma, the condition, below, should be provided \cite{tarey}.

\begin{equation}
 \gamma e^{\alpha d}=1
\end{equation}

In the case, $\gamma e^{\alpha d} \rightarrow 1$ ,the denominator of Equation 2.9 goes to zero and the number of electrons $N$ goes to infinity. Because of the situation, the electric breakdown occurs. 

In order to determine the value of $\alpha$, $\alpha$ should be written in terms of  parameters that change the value of $\alpha$. An expression, designated as Equation 2.12, shows  dependencies of $\alpha$:

\begin{equation}
 \alpha=A p e^{-B p/E}
\end{equation}
  
 In Equation 2.12, the constants $A$ and $B$  differ for different gases and $p$ is designating pressure of the gas. The  term $E$ determines the electric field of the inter-electrode space and therefore $E=V/d$ \cite{konuma}. The \textit{breakdown voltage} or \textit{starting voltage} $V_{br}$  can be computed by combining Equation 2.12 and Equation 2.10 \cite{konuma}. In addition to this, the breakdown is sustained at the room temperature $20^{o}$C and  the electron mobility is inversely proportional to pressure \cite{fridman}.

\begin{equation}
 A p d e^{-B p/E }=\ln{(1+\gamma^{-1})}
\end{equation}

By substituting the expression $E=V/d$ into  Equation 2.13, the \textit{breakdown voltage}, $V_{br}$, can be deduced as

\begin{equation}
 V_{br}=\frac{B p d}{\ln(p d)+\ln(A/\ln(1+1/\gamma))}
\end{equation}

\newpage
 Expression 2.14, is known as \textit{Paschen's Law} \cite{konuma,grill,tarey}.  '$pd$' is called as the \textit{reduced electrode distance}, on which the \textit{breakdown voltage} depends. The graph of  $V_{br}$ versus $pd$ is known as the \textit{Paschen curve}.  The value of $pd$ plays important role in the extraction process of ions. The typical \textit{Paschen curves} of several gases are given in Figure 2.2 \cite{lieberman2a}.

\begin{figure}[H]
 \centering
 \includegraphics[width=0.8\textwidth]{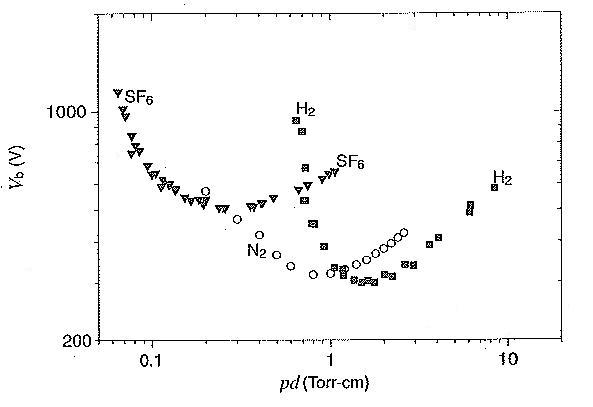}
 \caption{Paschen Curves at the temperature  $20^{o}$C [8](The graph is a log-log graph)}
\end{figure}

 A typical voltage which is needed to sustain the discharge depends upon the type of gas and the pressure used in the plasma chamber \cite{konuma}. And as it can be seen in Figure 2.2, the typical \textit{breakdown voltage} is at the order of hundreds of volts. However, there is  drastic increase at low pressures. Analysis for the low pressures gives the approximate values. The analysis is  done numerically by using \textit{GNU Octave}, which  is a scientific computing software. The value of $A$ and the value of $\ln(A/\ln(1+1/\gamma))$ are taken as 478.680 and 1.664 respectively. In order to compute the values of $A$ and the value of $\ln(A/\ln(1+1/\gamma))$, the data points (2.06,1800.00) and (11.00,259.09) are taken from Figure 2.3 \cite{lieberman1a} and   the estimated low pressure Paschen curve for the  hydrogen gas  is drawn as in Figure 2.4 . Note that the unit of $pd$ is in cm-torr in Figure 2.4.

\newpage
\begin{figure}[H]
 \centering
 \includegraphics[width=0.6\textwidth]{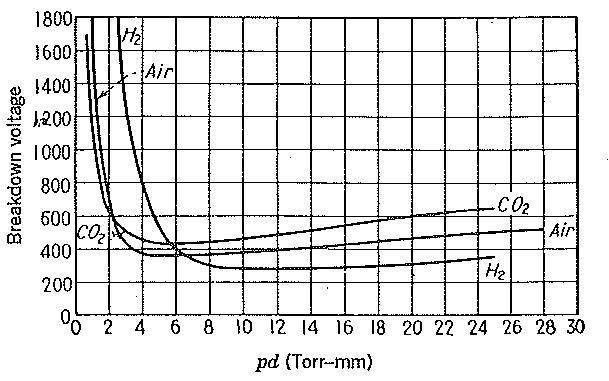}
 \caption{Paschen Curves at the temperature    $20^{o}$C [9]}
\end{figure}

\begin{figure}[H]
 \centering
 \includegraphics[width=0.7\textwidth]{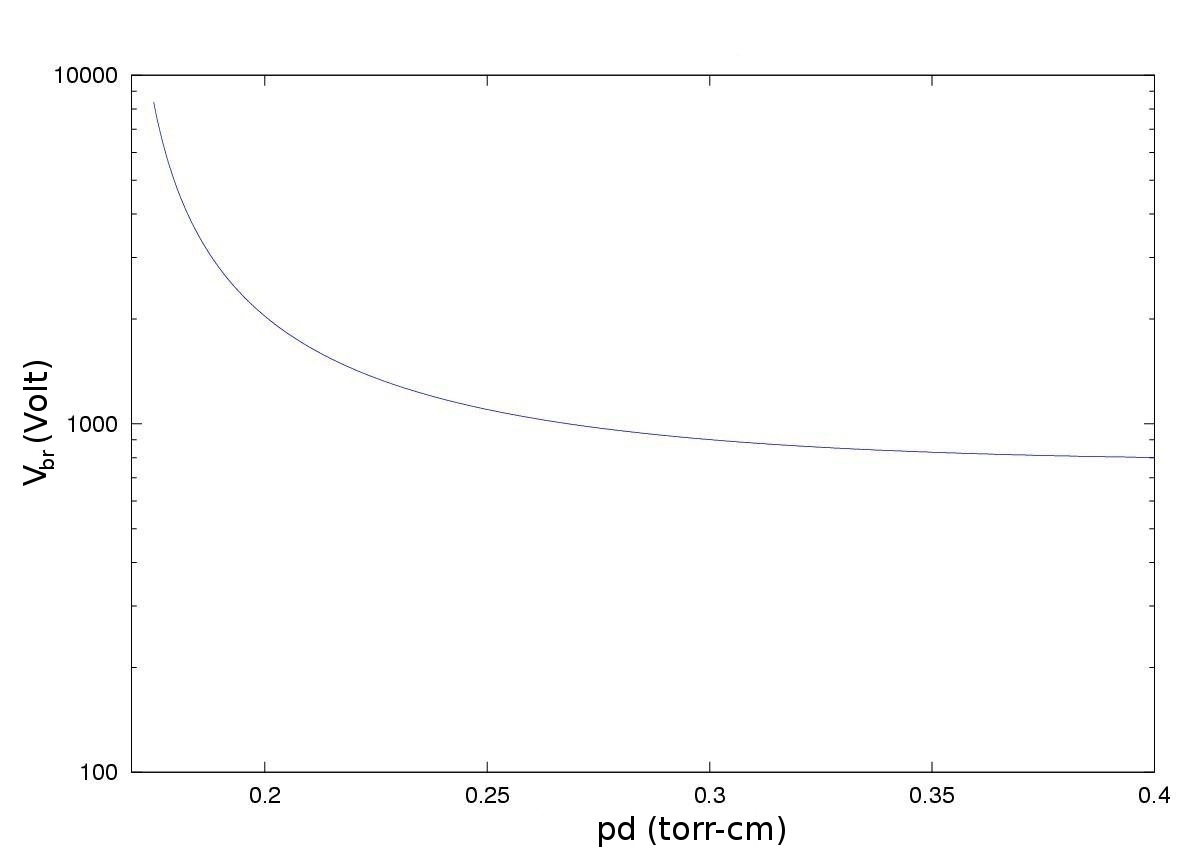}
 \caption{Estimated Paschen Curve for $H_{2}$. (This graph is a semilog graph.)}
\end{figure}

 The breakdown voltage for the low values of $pd$ is very important in order  to prevent the breakdown at the extraction zone  while applying high voltages. In addition to these, there is a crucial concept, \textit{plasma potential} which should be explained distinctly in order to understand the ion extraction process.

 DC discharge mechanism and DC discharge plasma ion sources are not the concern of  this thesis directly; however, in an ion extraction process, a DC plasma can occur. To prevent  the DC discharge at a vacuum gap, the mechanism of DC discharge should be well understood.

\section{MICROWAVE DISCHARGE}
% \subsection{MICROWAVE DISCHARGE} %referanslar: wolf2, pozar1, fridman1, fridman2, grill1, grill2

The microwave generated plasmas have similarities with the RF (Radio Frequency) generated plasmas and optically generated plasmas. The wavelength of the microwaves are at the range of centimeters, longer than the optical waves which are with  wavelengths in nanometer scale  and smaller than the wavelength of  RF radiation. Eventually, the all these  radiation types are ordinary electromagnetic radiations, so the plasma generation could be described by the basic electromagnetic theory. 

At first, the electric field of the microwave radiation can be considered in a rectangular waveguide. The microwaves can travel through any closed shaped metals. If these closed shaped metals are designed for the specific frequency, they guide  electromagnetic waves, passing through them. For the frequency of 2.45 GHz, the rectangular waveguides are used  and the mode of the microwave, which passes through them,  is $TE_{10}$ (Transverse Electric Field Mode 10, in rectangular waveguide).  The mode of the wave determines the electric fields and magnetic fields of the wave across the width and the height of the waveguide \cite{pozar1}. The electric field distribution  of the mode $TE_{10}$ is depicted as in Figure 2.5. The electric field has a maximum in the center of the width of the waveguide \cite{fridman}. The other modes can be used any microwave discharges but they are not the subject of this thesis. 

%2.5
\begin{figure}[H]
 \centering
 \includegraphics[width=0.5\textwidth]{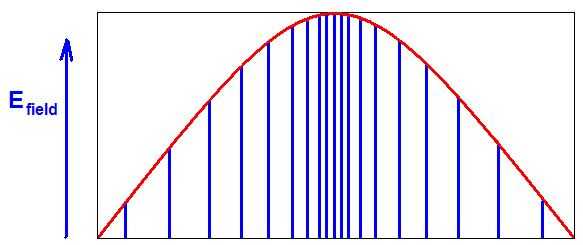}
 \caption{The $TE_{10}$ mode of the microwave in a rectangular waveguide}
\end{figure}

The free electrons, which are produced by the background UV radiation or gamma radiations, are made to oscillate by the alternating electric field of the microwave. The $TE_{10}$ mode is convenient for plasma generation in the rectangular waveguide because at the $TE_{10}$ mode, the electric field reaches its maximum value at the midpoint of the waveguide \cite{fridman}.   When the free electrons start  oscillating  with the oscillating electric field,  collisions with the neutral gas atoms   become  more violent, so these collisions heat up the gas. Due to the much larger mobility of the electrons with respect to ions , the heating process is accepted to be sustained by the electrons. The heavier ions can not respond the rapid change  in the electric field of the wave \cite{grill}. In this situation, the probability of  ionization  increases. 

%2.6
\begin{figure}[H]
 \centering
 \includegraphics[width=0.8\textwidth]{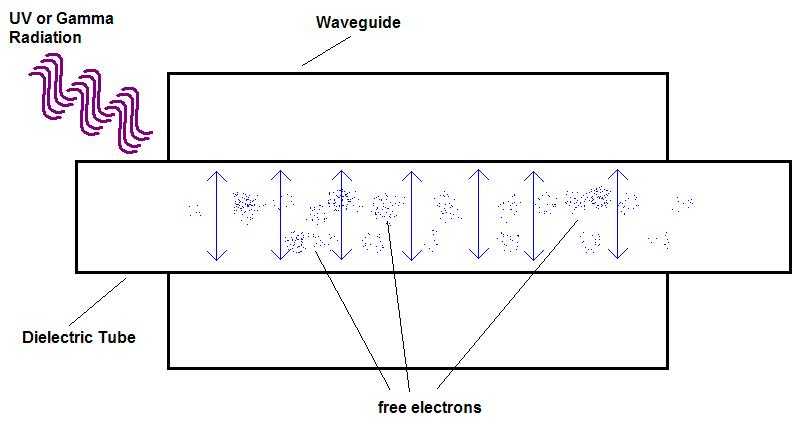}
 \caption{Ignition of a microwave plasma}
\end{figure}

The typical electric field strength of a 2.45 GHz microwave is approximately $ E_{0} \approx 30 V/cm $ \cite{grill}. If the collisionless situation( i.e., the electrons do not collide with ions or neutrals) is considered, the equations of the motion of the electrons can be derived as follows:

\begin{equation}
 \textbf{E}=E_{0}exp(-i \omega t)  \hat{\textbf{x}} \quad,
\end{equation}
where $E_{0}$ is a constant with the same dimensions as $\textbf{E}$.
\begin{equation*}
 \omega = 2\pi f \quad,
\end{equation*}

where  $\omega $ is the angular frequency of the wave and  $f$ is the frequency of a microwave source. In this work $f=2.45 GHz$.  

\begin{equation}
 m_{e}\ddot{\textbf{x}}=-e\textbf{E} \quad,
\end{equation}
where $e$ is elementary positive charge,
\begin{equation}
 \ddot{\textbf{x}}= -\frac{e}{m_{e}} E_{0}exp(-i \omega t)  \hat{\textbf{x}}
\end{equation}
by integrating the previous equation, we can get the velocity of the electron,

\begin{equation}
 \dot{\textbf{x}}= -\frac{eE_{0}}{m_{e} } \left( \frac{1}{-i\omega} \right) exp(-i \omega t)  \hat{\textbf{x}}
\end{equation}
and by the second integration, we can get the position of the electron,
 \begin{equation}
  \textbf{x}=\frac{eE_{0}}{m_{e} \omega^{2}} exp(-i \omega t)\hat{\textbf{x}}
 \end{equation}
The position of the electron is a function of time in Equation 2.19.  The position can be rewritten as follows (Note that the following equation is a scalar function!).
\begin{equation}
 x(t)=x_{0}exp(-i\omega t)
\end{equation}
The maximum distance $x_{0}$ that an electron is driven by the alternating electric field is given by
 \begin{equation}
  x_{0}=\frac{eE_{0}}{m_{e} \omega^{2}}
 \end{equation}

for the collisionless situation $ \nu / \omega<<1 $ should be provided where $\nu$ is the elastic collision frequency in gases.

The kinetic energy of the oscillating electron, $W_{e}$, is written as

\begin{equation}
 W_{e}=\frac{{m_{e}v_{e}^{2}}}{2} \quad,
\end{equation}
where $v_{e}$ is the speed of electron and it is the scalar quantity and the value of  $v_{e}$ is the magnitude of $\dot{\textbf{x}}$.
If the values of  a radiation with a frequency of 2.45 GHz  and a electric field strength of 30 V/cm  are substituted into the above equations by converting them to SI units, the maximum distance that the electrons travel $x<10^{-3}cm$ in a collisionless situation \cite{grill}. The maximum energy that an electron gains in one cycle is about 0.03 eV \cite{grill}. This amount of energy is not enough to ionize the neutral atoms. So if the ignition is to be made at low pressures, there should be some manipulations done to the experiment to increase the probability of the ionization.  

In collisional cases (the cases are at higher pressures than the first situations) it is to transfer the power from the outside electric field to the unit volume of gas and the average power $\bar{P}$ is given by \cite{grill}

\begin{equation}
 \bar{P}=\frac{n_{e}e^{2}E_{0}^{2}}{2m_{e}}           \left(\frac{m_{a}}{\nu^{2}+\omega^{2}}\right)  \quad,
\end{equation}

where $m_{a}$ is the mass of the atoms that the electrons collide with. The elastic collision frequency , $\nu$, is approximately in  the order of between $10^{9}$ and $10^{11}$ collisions per second at glow discharge conditions for gases \cite{grill}. 

The ionization of the neutral atoms becomes plausible in collisional situation.   Random motions of  electrons stem from the collisions with the atoms of the gas. Moreover, the electrons gain additional energy from the external electric field during  each collision with the atoms. After  an electron makes an elastic collision with an atom, it reverses its direction, at that instant of time, if the electric field  changes direction,  the electron accumulate enough energy to ionize the neutral atoms \cite{grill}. 

%\newpage
The plasma is ignited and maintained by the absorption of the electromagnetic energy and the heat exchange of the thermal plasma is provided by the convective cooling in the gas flow \cite{fridman}.

\begin{figure}[H]
 \centering
 \includegraphics[width=0.8\textwidth]{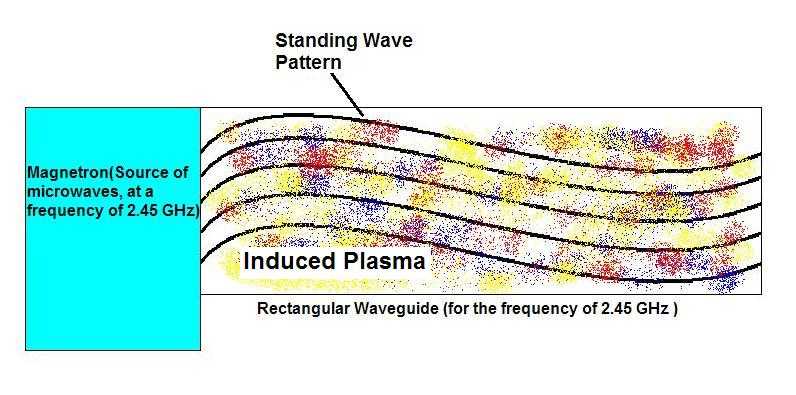}
 \caption{Standing wave pattern within the induced plasma}
 \label{Standing wave pattern within the induced plasma}
\end{figure}

\begin{figure}[H]
 \centering
 \includegraphics[width=1\textwidth]{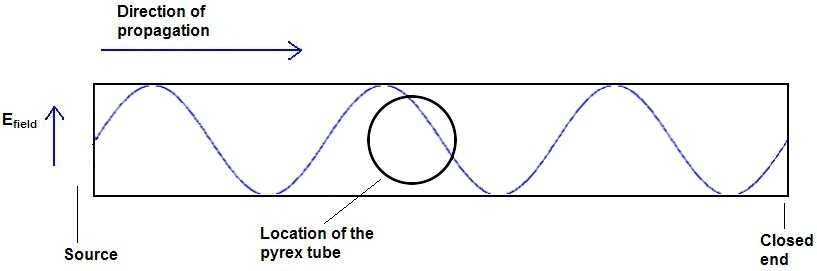}
 \caption{Location of the plasma tube across the standing wave}
 \label{Location of the plasma tube across the standing wave}
\end{figure}

The incident electromagnetic radiation can only transfer a certain amount of its energy  while traveling through the waveguide. The remaining  energy is transmitted forward or reflected backward to the source by the thermal plasma. In order to increase the dissipated energy by the thermal plasma, the standing wave  is formed by trapping the electromagnetic wave (i.e.,  the microwave.) \cite{fridman}. As it is shown in Figure \ref{Standing wave pattern within the induced plasma} and also in Figure \ref{Location of the plasma tube across the standing wave}, the standing wave pattern is formed through  the waveguide. If special equipments like  a stub tuner, a plunger, an isolator and a directional coupler, are used, the fraction of absorbed electromagnetic energy  can be increased up to 90\% to 95\% \cite{fridman}.

%2.7

For the our designed system the location of the pyrex plasma discharge tube, which is almost transparent for the microwaves (instead of the pyrex, quartz is usually used), crosses the waveguide horizontally.  

%2.8

The microwave has a wavelength of 12.24 cm and the length of the waveguide is  three times the wavelength of the corresponding microwave. There is a standing wave pattern of the system in Figure \ref{Location of the plasma tube across the standing wave}. The source of the microwave is on the left side of the waveguide and the right side of the waveguide is totally closed and sealed. The pyrex tube is intersecting the waveguide horizontally between the node and the antinode of the standing wave. At the left end of Figure \ref{Location of the plasma tube across the standing wave}, the oscillation of the electric field is shown. The oscillations of the electric field in the form of the standing wave induce a plasma inside the pyrex tube as  shown.

Before, skipping to a new chapter, there is a very crucial remark that in the absence of magnetic confinement for  the microwave discharge plasma. There is an  upper limit for the electron number  density (i.e., \textit{number of electrons per cubic centimeter}),  $n_{e}$, in the plasma \cite{wolf}.

\begin{equation}
 n_{e}\leq 1.11 \times 10^{10} f^{2} \quad,
\end{equation}

where $f$ is the frequency in GHz, and for the microwave with the frequency of 2.45 GHz the upper limit of the electron density is   $n_{e}=6.66 \times 10^{10} cm^{-3}$ \cite{wolf}. The reason of existence of  the upper limit is that the  microwave will not be able to penetrate  the plasma without applying a sufficiently high magnetic field (for $f= 2.45$ GHz, $B = 0.0875$ Tesla) through the plasma when the electron density reaches the upper limit.  Moreover, the power of the microwave is absorbed at the surface of the plasma and the most  of it reflects \cite{wolf}.

The details of the designed and built system are described in the following chapters with the help of the theoretical explanations,  mentioned above.

% \section{PLASMA DIAGNOSTIC BY DOUBLE LANGMUIR PROBE} bu bölümü chapter 5 e dahil et.

\newpage

% CHAPTER 3
\chapter{EXTRACTION OF IONS FROM PLASMA}
\label{chp:extraction of ions from plasma}
 The produced ions should be extracted from the plasma medium in order to direct them to targets or guide them to ion beam lines to distort  profiles of ion beams. In order to collect the ions, a positive electrical potential is applied to the plasma medium with respect to the extracted zone and then  ions are ready to be accelerated at  the edge of the plasma medium. Accelerated ions are able to be guided and directed by the electric field  and also a dipole magnet. In order to understand this procedure, \textit{plasma (space) potential,  extraction } should be discussed firstly. Also, the use of \textit{vacuum  apertures} will be explained for the production of \textit{vacuum gradient}.

\section{PLASMA (SPACE) POTENTIAL} % referanslar: chenc,fridmanc,konumad,noyesa,browna 
 For any practical plasma devices, plasma and  walls of the plasma chamber produce a region, called sheath. Consider   a plasma system  which has no electric field inside or  applied by any electrodes, inserted in. Since the electrons move much faster than the ions in the plasma, the electrons leave the plasma  at a greater rate than the ions do \cite{chen,fridman,brown}.  The thermal velocities of the individual particles can be calculated for the electrons and ions by using  Equation 3.1 and 3.2 respectively. 
 
\begin{equation}
 v_{e}=\sqrt{\frac{3kT_{e}}{m_{e}}}  \quad,
\end{equation}

\begin{equation}
 v_{i}=\sqrt{\frac{3kT_{i}}{M_{i}}} \quad,
\end{equation}
where $v_{e}$ is the speed of an electron and $v_{i}$ is the speed of an ion. In Equation 3.2, the temperature of ions $T_{i}$ and   the mass of an ion  $M_{i}$  should be considered. The value of $M_{i}/m_{e}$ is about 1837 for a hydrogen ion. From these  equations, it can be deduced that the electrons' thermal velocities are about 1000 times faster than the ions' thermal velocities \cite{fridman}. The fast moving electrons leave the plasma and stick to the walls of the chamber and  the slow ions are assumed to be stationary with respect to the electrons, so that the plasma remains with a net positive charge \cite{chen,fridman}. These excess charges produce a potential with respect to the walls.  Debye shielding forms a potential variation  in several Debye lengths. This layer is called \textit{plasma sheath} \cite{chen,brown}. As F.F. Chen mentioned ``The function of a sheath is to form  a potential barier so that the more mobile species, usually electrons, is confined electrostatically. The height of the barrier adjust itself so that the flux of electrons that have enough energy to go over the barrier to the wall is just equal to the flux of ions reaching the wall \cite{chen}''. The plasma potential drops while approaching the wall. An illustration is shown in Figure 3.1.

\begin{figure}[H]
 \centering
 \includegraphics[width=0.8\textwidth]{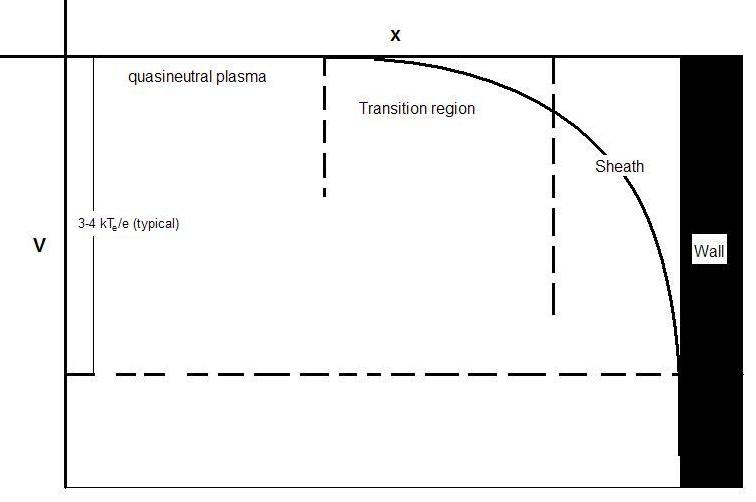}
 \caption{Plasma potential and the regions between the plasma and the wall}
\end{figure}  

 The reduction of the potential depends on the configuration and  the plasma parameters yet it is typically 3 or 4 times the electron temperature (in electron volts) per  charge of an electron (in Coulomb) and this situation is also valid for a floating probe or electrode which inserted in a plasma \cite{brown}. The floating electrode or probe can be said to be at the floating potential, negative with respect to the plasma potential.

 In a situation that a conducting electrode immersed near the sheath and is held at a lower potential with respect to the plasma,  the electrode attracts the ions and repels the electrons. If the electron density is taken to be zero at the sheath, the current density $J$ can be written as  \cite{noyes}. Note that the charge state (or the ionization state) is taken as 1 for convenience while doing the derivations.

\begin{equation}
 J=n_{i} e v_{i} \quad,
\end{equation}
 
 where $n_{i}$ is the number density,  $e$ is the elementary charge in Coulombs and $v_{i}$ is the speed of ions, respectively.   The Poisson equation is written for this region as

\begin{equation}
 \frac{d^{2}V}{ dx^{2} }=\frac{ n_{i} e}{ \varepsilon_{0} }
\end{equation}

 By the conservation of energy (The calculations are non-relativistic calculations!), 

\begin{equation}
 M_{i}v_{i}^{2}/2=eV \quad,
\end{equation}

where $M_{i}$ is the individual mass of the ions. The term $ n_{i} e$ can be written in terms of $J$ and $v_{i}$, then the Poisson equation becomes

\begin{equation}
 \frac{d^{2}V}{dx^{2}}=\frac{1}{\varepsilon_{0}}\sqrt{\frac{M_{i}}{2e}}\sqrt{\frac{1}{V}} J(x)
\end{equation}

If Equation 3.6 is multiplied by $(dV/dx)$ and integrated from zero to a potential $V$ \cite {noyes,diag}, we get

\begin{equation}
 \left(\frac{dV}{dx}\right)^{2}=\frac{4}{\varepsilon_{0}}\sqrt{\frac{M_{i}}{2e}}\sqrt{\frac{1}{V}} J(x)
\end{equation}

 By taking the square root of both sides of Equation 3.7, and then rearranging it,

\begin{equation}
 \frac{dV}{V^{1/4}}=\frac{2}{\varepsilon_{0}}\sqrt{\frac{M_{i}}{2e}}J^{1/2}(x)dx
\end{equation}
 and then by integrating the above equation, the current density is found as

\begin{equation}
 J(x)=\frac{4\varepsilon_{0}}{9}\sqrt{\frac{2e}{M_{i}}}\frac{V^{3/2}}{x^{2}}
\end{equation}
 
 Equation 3.9 is known as Child-Langmuir Law of space charge-limited current flow \cite{chen,noyes,diag} and more general notation for ions having charge state (or ionization state) $Z$ is \cite{roth}

\begin{equation}
 J(x)=\frac{4\varepsilon_{0}}{9}\sqrt{\frac{2eZ}{M_{i}}}\frac{V_{0}^{3/2}}{x^{2}}   
\end{equation}
 
The Debye length is the crucial for the boundary of the plasma and for the dynamics of the transition region. For example, in the case of the high voltage  applied to a electrode inside the plasma, the sheath region gets thicker than the Debye length and the sheath distance $d_{sheath}$ can be written approximately as   \cite{brown}.

\begin{equation}
 d_{sheath}\sim \lambda_{D}\sqrt{\frac{eV}{kT_{e}}}
\end{equation}
 
Equation 3.11 becomes very important in order to determine the probe sizes, dimension of wire meshes and the extraction apertures and separation \cite{brown}.

\section{EXTRACTION} %referanslar: brownb,  rothe, wolfd
The extraction process of ions from the generated plasma plays a  crucial role while developing an ion source. Ions move randomly  inside the plasma with free electrons. Each  ion  has a different kinetic energy. The ions should be made mono energetic and unidirectional before they are injected to a process chamber \cite{roth}. This is the reason why multifarious electrodes are used. The general types of these are named as \textit{accelerating grids or extractor grids or extractor mesh, (axial) diode system, triode diode extractor, mutigap extraction systems} and \textit{ multiaperture extractors} \cite{wolf,brown}. All these types  are basically simple particle accelerators with special geometrical design. The extractors should be designed while  the limits of the Child-Langmuir law is being taken into consideration. 

\begin{figure}[h]
 \centering
 \includegraphics[width=0.7\textwidth]{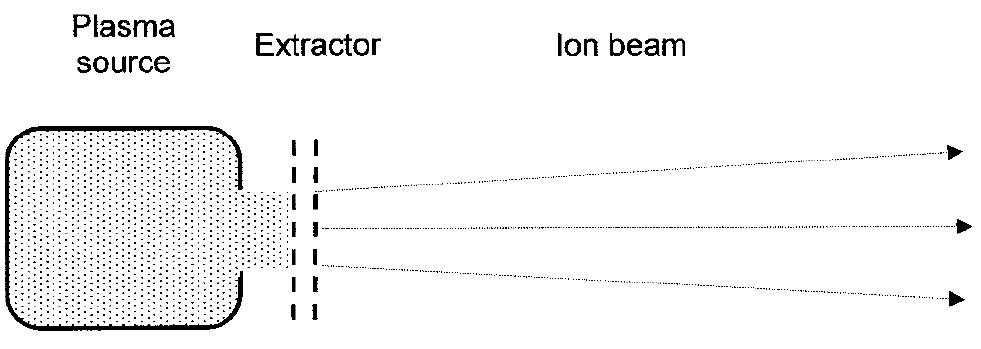}
 \caption{A basic extraction system with  mesh extractors}
 \label{A basic extraction system with  mesh extractors}
\end{figure}

\begin{figure}[h]
 \centering
 \includegraphics[width=0.8\textwidth]{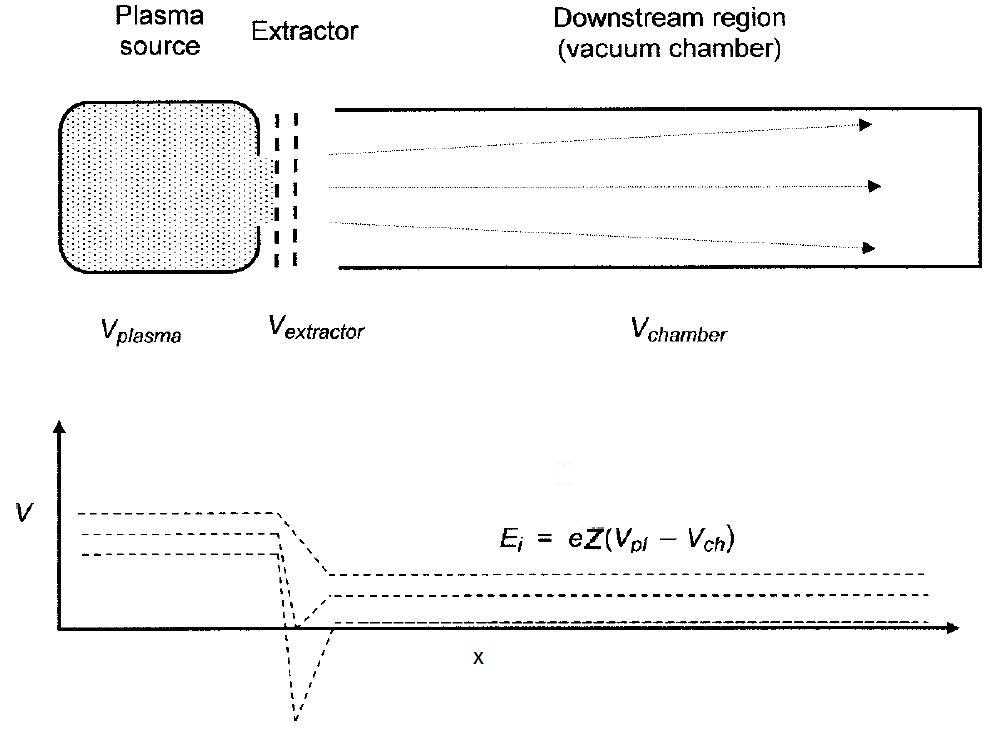}
 \caption{Extraction with the different extraction voltages}
 \label{Extraction with the different extraction voltages}
\end{figure}

\begin{figure}[H]
 \centering
 \includegraphics[width=0.48\textwidth]{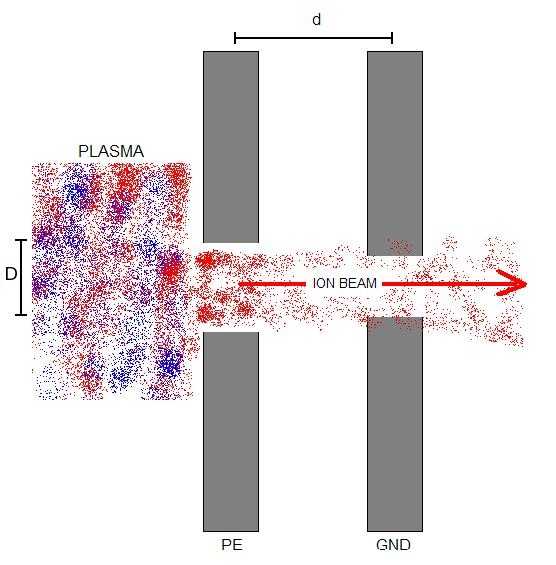}
 \caption{Extraction aperture diagram}
 \label{Extraction aperture diagram}
\end{figure}

Typical extraction techniques   are shown in figures \ref{A basic extraction system with  mesh extractors} , \ref{Extraction with the different extraction voltages} , \ref{Extraction aperture diagram} \cite{brown}. In the present work, the technique shown in Figure \ref{Extraction aperture diagram} is modified. A basic extraction system can be designed as in Figure \ref{A basic extraction system with  mesh extractors}. For this design the target of the ion beams are not defined. For  usual purposes, the ions are driven to  a vacuum chamber (or downstream region). This vacuum chamber can be held at a constant electrical potential. For the general case, if the electrical potential is applied to the vacuum chamber, where the ions 
%3.2 dikkat bu figure n yerien h konumunda kontrolet!!!!!!

impinge the corresponding target, the potential of the downstream region (vacuum chamber) should be taken into consideration. If the plasma is at the potential, $V_{pl}$, and the extractors' voltages are at the values, $V_{ext1}$, $V_{ext2}$, $V_{ext3}$..., and the potential of the vacuum chamber is at the potential,  $V_{ch}$, then the general case of a extraction system can be thought as in Figure \ref{Extraction with the different extraction voltages} with the given voltages.

%3.3

If the charge state  $Z$ of the ions are known,  the energy of the ions, at the downstream region, can be estimated as

\begin{equation}
 E_{i}=eZ(V_{pl}-V_{ch}) \quad,
\end{equation}

where $e$ is the electron charge in units of Coulomb. Here the plasma potential is increased with respect to the chamber potential or any electrodes' potential by applying bias voltage to the plasma chamber.  Nevertheless, the plasma voltage is higher than any potential, applied to the plasma chamber or any electrodes which are immersed  in the plasma. The  plasma potential is  $3kT_{e}/e$ volts or $4kT_{e}/e$ volts higher than any high voltage electrode or frame, touching the plasma. It can be a good approximation when the plasma potential, $V_{pl}$, is assumed to be  equal to the potential of a  extractor, $V_{ext}$,  which are typically at the order of kilovolts \cite{brown}.  For  low energy ion sources and their applications, the plasma potential should be added to the voltage of the extractor and should be taken into consideration \cite{brown}. 

In this chapter, the axial diode extractor system, with a single axial aperture, is going to be discussed. For this type of extractor,  a single aperture is used in order to let the ions go through it with the diameter, $D$, and the second aperture is located at a distance, $d$. Both apertures can have the same diameter. For the general case, the diameters of the two apertures are different as they are shown in Figure \ref{Extraction aperture diagram}. 

%3.4

For the designed system, the second aperture is narrower  than the first single aperture, in order to maintain the vacuum gradient between them. The vacuum gradient is discussed in the following section.  It is worth to mention that the second aperture (at the second grounded electrode) determines the initial diameter of the ion beam. Plasma potential or the potential of the electrode, the neighbouring electrode of the plasma, is labeled as PE and the distant electrode is labeled as GND which means the \textit{grounded electrode} in Figure \ref{Extraction aperture diagram}.

\begin{figure}[H]
 \centering
 \includegraphics[width=0.8\textwidth]{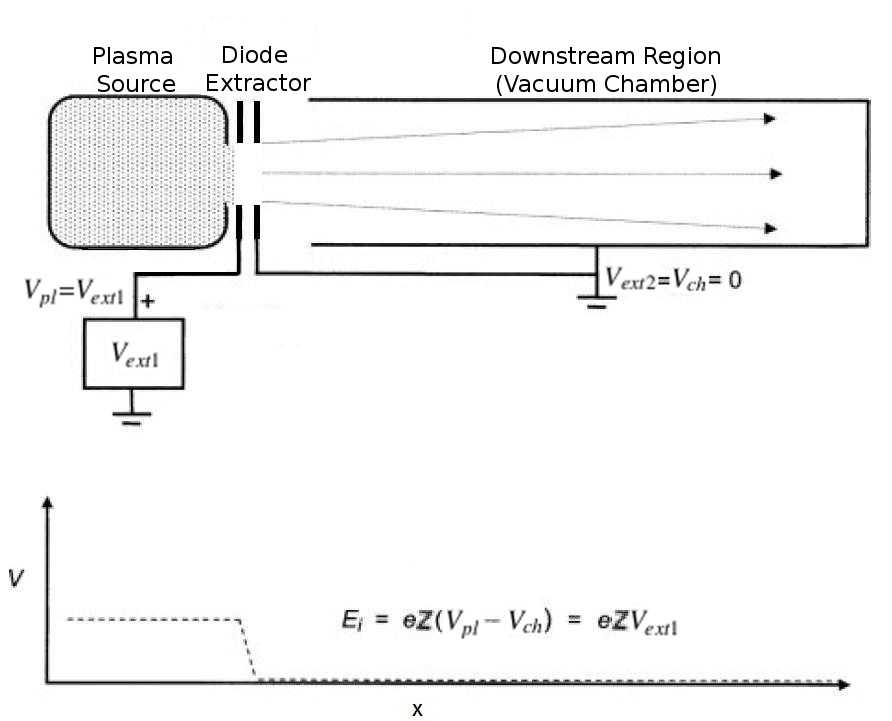}
 \caption{The extraction diagram for the designed system}
 \label{The extraction diagram for the designed system}
\end{figure}

This diode extractor is placed in front of the plasma source.  It is important to be near the plasma. This extraction system is located as  shown in Figure \ref{The extraction diagram for the designed system}.  The vacuum chamber is located after the diode extractor. A high voltage (about 3 kV) is applied to the first aperture electrode and the second aperture electrode and the vacuum chamber are grounded as they are shown in Figure \ref{The extraction diagram for the designed system}.  The potential of the plasma, $V_{pl}$, is assumed approximately  equal to the potential of the first extractor, $V_{ext1}$  $(V_{pl}=V_{ext1})$.  Both the potential of the second extractor, $V_{ext2}$, and the potential of the vacuum chamber, $V_{ch}$,  are grounded. Then the system becomes as it is shown in Figure 3.5. The energy of the  beam ions is given by,
\begin{equation}
 E_{i}=eZ V_{ext1}
\end{equation}

% figure mesh değil a single aperture yap
%3.5

Figure \ref{The extraction diagram for the designed system} is the schematic diagram of the designed system. For  the rest of the thesis, discussions would be based on this configuration. The importance of  ion extraction will be discussed in Chapter 4.

\section{VACUUM GRADIENT}
\label{sec:VACUUM GRADIENT}
Vacuum process is very important in this system. Unless the plasma is generated at a specific pressure, it will not be sustained and will not be stable. Moreover, the extraction  process of the ions  should be succeeded at very low pressures in order to  prevent the DC discharge between the extraction electrodes and the metal frames of the system. 

\begin{figure}[H]
 \centering
 \includegraphics[width=0.8\textwidth]{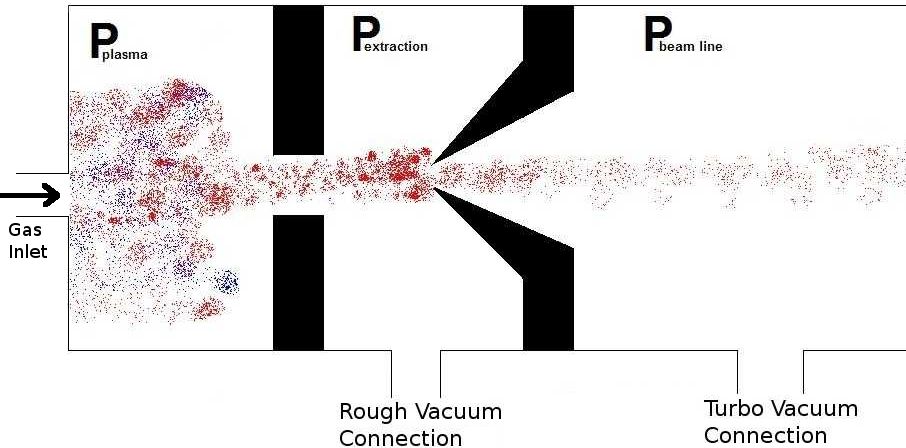}
 \caption{Vacuum zones in order to form the vacuum gradient}
 \label{Vacuum zones in order to form the vacuum gradient}
\end{figure}

In the designed system, the plasma chamber (tube) is held at a pressure of $P_{plasma}$. This pressure defines the plasma pressure. The place  where the extraction electrodes are separated, is held at a pressure of $P_{extraction}$ which is lower than the plasma pressure, $P_{plasma}$.  Finally, the ion beam line or the measurement vacuum gradient part ends with a place which is held at the lowest pressure,  named as  $P_{beam line}$. The scheme is shown in Figure \ref{Vacuum zones in order to form the vacuum gradient}. The plasma pressure   is controlled by the gas inlet  which is located at the left side of the figure.   The gap  between the extraction electrodes  is vacuumed by the rough vacuum. The rough vacuum of the whole system  is  also made from this connection as it is shown in Figure \ref{Vacuum zones in order to form the vacuum gradient}. The  system is brought to the highest vacuum levels by the high vacuum pumps through the turbo vacuum connection.

%3.6

\newpage
Finally,  by applying the vacuuming process part by part, the vacuum gradient is sustained. To maintain the vacuum gradient, the separation of the apertures  and   the extraction aperture should be small enough and also the geometry of the apertures is very important.  The conical separator is used to achieve the vacuum gradient in the  system. In addition to these, the number of the vacuum separators  between the vacuumed parts is also significant.

\newpage

% CHAPTER 4
\chapter{CHARACTERIZATION OF ION BEAM AND ION SOURCE} %referanslar: brownc, browncc, rothb,rothc, rothd, wolfe, wolff
\label{chp:characterization of ion beam and ion source}

\section{PERVEANCE}%referanslar: browncc, rothb, wolfe
 Perveance defines the relation between the electron or ion beam current and the extraction voltage of the source with fixed cross-sectional beam area \cite{roth}. The perveance is given as
 
\begin{equation}
 \mathcal{P}\equiv \frac{I_{b}}{V_{0}^{3/2}}=\frac{J \pi D^{2}}{4 V_{0}^{3/2}}   
\end{equation}

 In the above equation, $I_{b}$ is the beam current and  $V_{0}$ is the extraction voltage. $D$ is defined as the initial beam diameter whereas $J$ is the density of the beam current (look at Figure 3.4).
 For  ideal ion sources, the extraction voltage would be applied via planar accelerating grids separated by a distance $d$. The extraction voltage $V_{0}$ would operate with the accelerating grid  with no webbing \cite{roth}. The ion current density can be written via the \textit{Child space-charge limited condition}.

\begin{equation}
 J=\frac{4\varepsilon_{0}}{9}\sqrt{\frac{2eZ}{M_{i}}}\frac{V_{0}^{3/2}}{d^{2}}   
\end{equation}

Here,  $Z$ indicates the charge state (or ionization state) of the ions.
The Child-Langmuir law perfectly suits for estimating the maximum value of  ion beam current \cite{brown}.  More practical form of Equation 4.2 is given by \cite{wolf}
 
\begin{equation}
 J=(1.72) \sqrt{\frac{Z}{u}}\frac{V^{3/2}}{d^{2}}
\end{equation}

where $J$ is in  units of milliampere per square centimeter  $(mA/cm^{2})$, $u$ is the atomic mass unit of the ions, $V$ is in  units of kilovolt $(kV)$, and $d$ is in the unit of millimeter $(mm)$.

The ion  beam current  $I_{b}$, is determined by the smallest aperture of the extraction electrode and by multiplying the current density   $J$, with the area of the aperture, $\pi D^{2}/4$. The net current is found  as

\begin{equation}
I_{b}=\frac{\pi D^{2}}{4} J
\end{equation}

\begin{equation}
 I_{b}=\frac{\varepsilon_{0} \pi }{9}\sqrt{\frac{2eZ}{M_{i}}}\frac{V_{0}^{3/2} D^{2}}{d^{2}}
\end{equation}

For an ideal ion source the perveance is found by substituting Equation 4.2 into Equation 4.1 as,

\begin{equation}
 \mathcal{P}_{max} = \frac{J \pi D^{2}}{4V_{0}^{3/2}}=\frac{ \pi \varepsilon_{0}}{9} \sqrt{ \frac{2eZ}{M_{i}}} \left(\frac{D}{d} \right)^2
\end{equation}

 It is better  to design an ion source, having the value of the perveance which should be close the theoretical value of the perveance $ \mathcal{P}_{max} $. In this work, the theoretical value of $ \mathcal{P}_{max} $ is calculated as $2.53\times 10^{-10} (A/V^{3/2})$ with respect to the  dimensions of the designed extraction unit.

%\section{ENERGY CONSUMPTION}
 %The required power is important to 

\section{EFFICIENCY OF ION SOURCES}
The power consumption of an ion source $P_{is}$, can be written by the sum of the power consumption of the extraction  unit $P_{eu}$, and the power consumption of the plasma generator $P_{pg}$.

\begin{equation}
 P_{is}=P_{eu}+P_{pg}
\end{equation}
 the power consumption of the extraction unit is given by \cite{roth},

\begin{equation}
 P_{eu}=I_{b}V_{0}
\end{equation}

So Equation 4.7 becomes 

\begin{equation}
 P_{is}=I_{b}(V_{0}+V')    \quad,
\end{equation}
where $V'$ is the required potential $(V)$ by the source to produce the ions and $I_{b}$ is the ion current \cite{roth}. $V'$ can be obtained by measuring the power consumption of an ion production unit which is used to produce raw ions for the system in order to obtain the ion current $I_{b}$. The power consumption of the ion production unit is calculated by $W_{ipu}=I_{b}V'$ for  electrical devices. Where $W_{ipu}$ is the  dissipated  power by the ion production unit for the corresponding ion current $I_{b}$.
The electrical efficiency of the ion  source, $\eta_{e}$, can be written as 

\begin{equation}
 \eta_{e} =\frac{P_{eu}}{P_{is}}=\frac{V_{0}}{V_{0}+V'}
\end{equation}
\begin{center}
or alternatively,
\end{center}
\begin{equation}
 \eta_{e} =\frac{1}{1+(V'/V_{0})}
\end{equation}

\section{REQUIRED GAS FLOW RATE}

The imperative ingredient to provide the required plasma density  is the gas . Firstly, the ratio of the ions' out flow   from the source to the in-flow of the neutrals into the source determines the the gas utilization efficiency, $\eta_{g}$ \cite{roth}.

\begin{equation}
 \eta_{g}= \frac{S_{i}}{S_{n}}=\frac{I_{b}}{eS_{n}} \quad,
\end{equation}

where $S_{i}$ and $S_{n}$ are the out-flow and the in-flow rates, respectively. The term, $S_{n}$, is the flow rate of the neutrals into the plasma chamber.

The net efficiency of the ion source $\eta_{is}$ is calculated by multiplying the gas utilization efficiency  $\eta_{g}$, with the electrical efficiency  $\eta_{e}$ \cite{roth}.

\begin{equation}
 \eta_{is}=\eta_{g} \eta_{e}
\end{equation}
If the in-flow rate of the gas is too low, it causes  an  extra load to the plasma generator and the efficiency  decreases. The good utilization of the gas, for supplying the propellant to the ion source, is a very important effect  on the ion sources. 

\section{ION FLUX AND NUMBER DENSITY OF ION BEAM} %reference: rothd
In order to find out the number of the particles passing through   a region per unit time, the ion flux should be considered. The ion flux, $\phi_{i}$, of the system can be written by using the initial beam diameter, $D$ \cite{roth}. 
\begin{equation}
 \phi_{i}= \frac{J}{eZ} 
\end{equation}
\begin{center}
 and
\end{center}
\begin{equation}
 J=\frac{4I_{b}}{\pi D^{2} } 
\end{equation}
by substituting Equation 4.15   into  Equation 4.14
\begin{equation}
 \phi_{i}= \frac{4 I_{b}}{\pi e Z D^{2}}
\end{equation}

$\phi_{i}$ is in units of ions per square meter per second $(ions/m^{2}s)$. Moreover, the number density of the ion beam is calculated by using  Equation 4.4 \cite{roth}, and rewriting the equation  for $n_{b}$ which is the number density of the ion beam:

\begin{equation}
 J=Ze n_{b} v_{b} \quad,
\end{equation}

\begin{equation}
 n_{b} =\frac{J}{Ze v_{b}}
\end{equation}
\begin{center}
 or
\end{center}
\begin{equation}
 n_{b} =\frac{\phi_{i}}{v_{b}} \quad,
\end{equation}

where $v_{b}$ is the mean velocity of the ions for the applied extraction voltage, $V_{0}$. From the conservation of energy  (here the thermal energy of the ions are neglected) the mean velocity, $v_{b}$, is written  as

\begin{equation}
 v_{b}=\sqrt{\frac{M_{i}}{2eZ V_{0}}}
\end{equation}

If equations 4.2 and 4.20 are substituted into  Equation 4.18, the number density of the ion beam, $n_{b}$, is found as

\begin{equation}
 n_{b} =\frac{4\varepsilon_{0} V_{0}}{9eZ d^{2}}
\end{equation}

As it can be seen explicitly in  Equation 4.21, the ion density is independent of the ion mass and the unit of the number density is in  ions per square meter  $(ions/m^{2})$.

\newpage

% CHAPTER 5
\chapter{THE DESIGNED SYSTEM}
\label{chp:the designed system}
 A  microwave coupled plasma ion source is designed and constructed  in this work. The microwave generation is provided by a magnetron which has the output power of 700 Watt with the frequency of 2.45 GHz. Since the high voltage power supply of the magnetron alternates with the frequency of 50 Hz,  the generated microwaves pulse with the frequency of 50 Hz. The power supply of the magnetron has a half-wave rectifier, so the magnetron also turns off after each cycle of the alternating current.    When the magnetron is turned on, the microwave generation starts   with the maximum power. The magnetron is cooled by  air ventilation. The \textit{3D design} of the microwave generator is shown in Figure \ref{Microwave generator with the integrated power source, the waveguide and a pyrex plasma chamber}.

%5.1
\begin{figure}[H]
 \centering
 \includegraphics[width=0.6\textwidth]{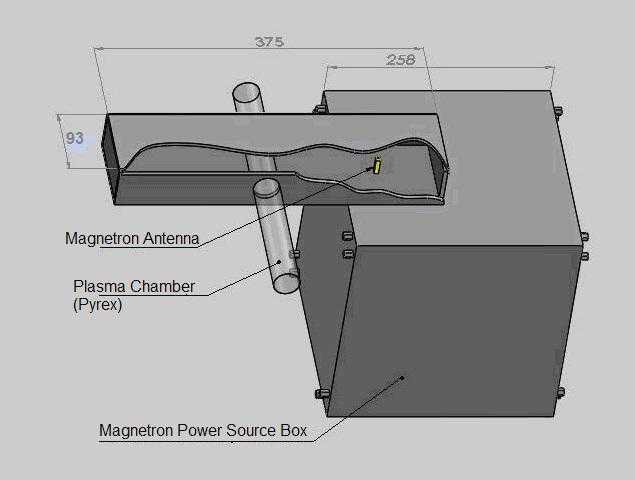}
 \caption{Microwave generator with the integrated power source, the waveguide and a pyrex plasma chamber}
 \label{Microwave generator with the integrated power source, the waveguide and a pyrex plasma chamber}
\end{figure} 

 The dimensions of the waveguide are determined with respect to electromagnetic waves with the frequency of 2.45 GHz which sits in the microwave region of the electromagnetic spectrum. The inside height of waveguide is 45 mm whereas the inside width  is 93 mm. Because of the type of the rectangular waveguide, the mode of the propagating wave is $TE_{10}$ inside the waveguide. 

\begin{figure}[H]
 \centering
 \includegraphics[width=1\textwidth]{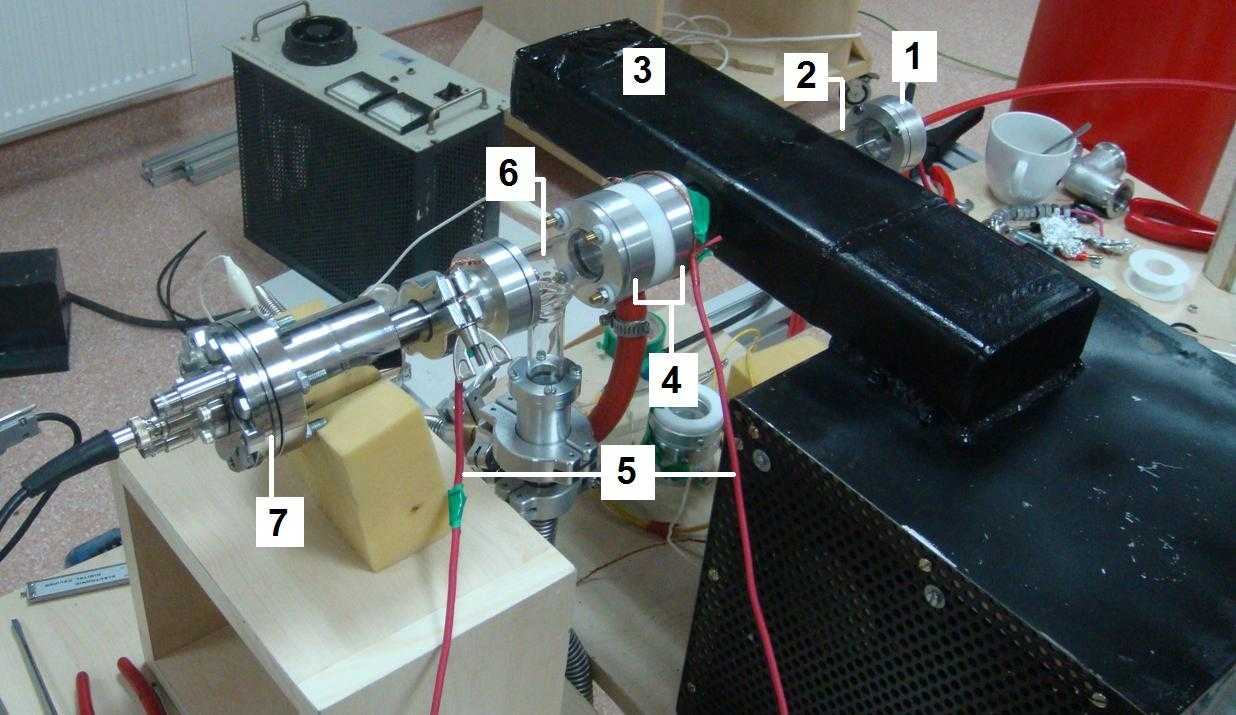}
 \caption{Microwave generator with the integrated power source and the general view of the assembled system}
 \label{Microwave generator with the integrated power source and the general view of the assembled system}
\end{figure}

 As they are shown in Figure \ref{Microwave generator with the integrated power source and the general view of the assembled system}, all the parts of the system numbered and titles of the numbered parts are given in Table \ref{Parts and components of the system}. 

%5.2 % generalviewsmallscalenumbered numaraları beya fon koyarak arkasına paint te düzelt ve jepg yap.

\begin{table}[H]
\caption{Parts and components of the system}
  \begin{center}
    \begin{tabular}{| l|c |}
    \hline % horizontal line demek
    1 & Gas Inlet \\
    \hline
    2 & Plasma Chamber (Pyrex Tube) \\
    \hline
    3 & Waveguide \\
    \hline
    4 & Extraction Electrodes \\
    \hline
    5 & High Voltage Cables \\
    \hline
    6 & Vacuum Gradient \\
    \hline
    7 & Faraday Cup \\
    \hline
    \end{tabular}
  \end{center}
  
  \label{Parts and components of the system}
\end{table}

\begin{figure}[H]
 \centering
 \includegraphics[width=1\textwidth]{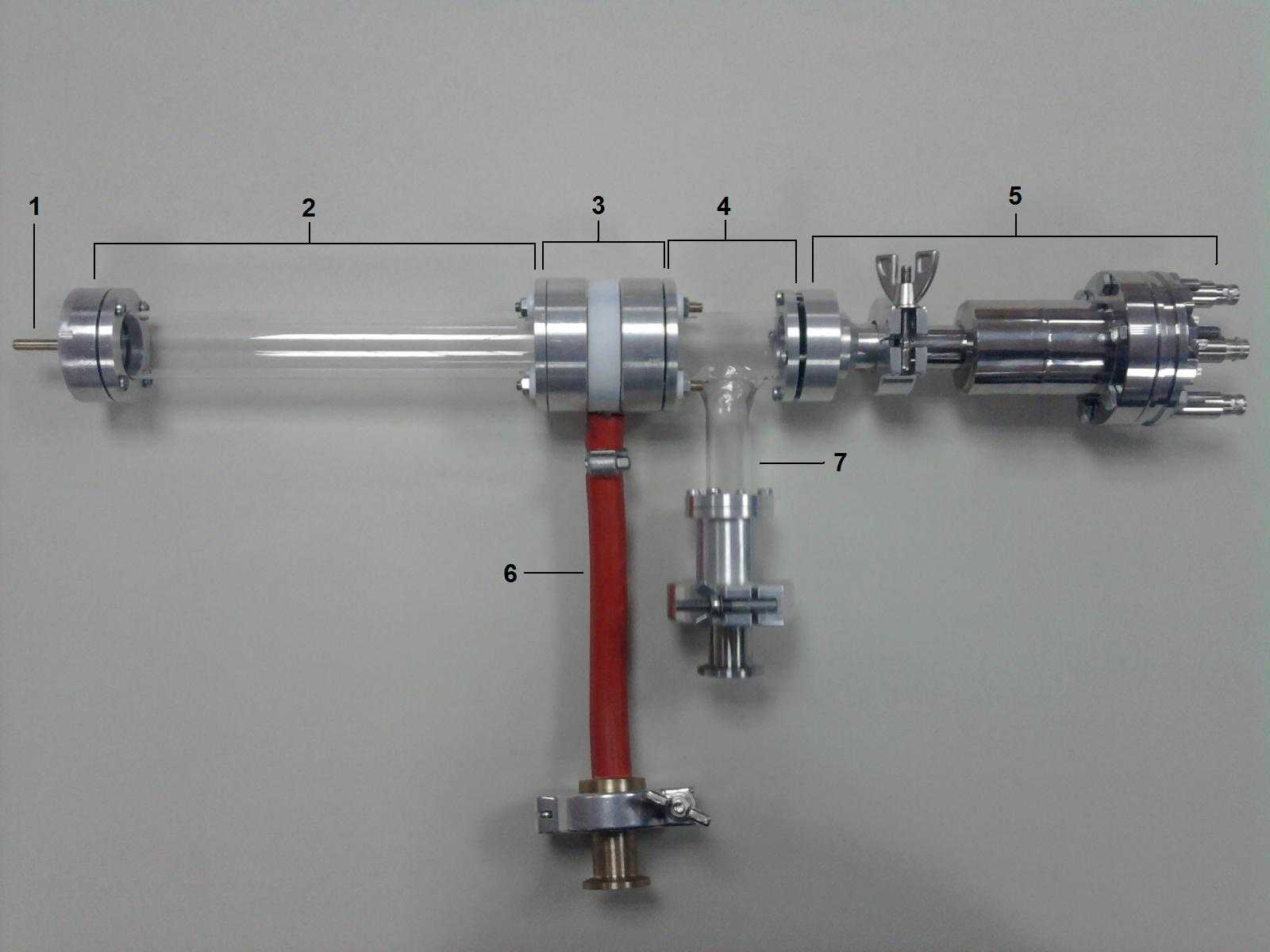}
 \caption{Assembled system without the microwave source}
 \label{Assembled System without the microwave source}
\end{figure}

 In Figure \ref{Assembled System without the microwave source}, the system is unmounted from the waveguide and the parts are named as in Table \ref{Names of the parts in Figure 5.3}.
%5.3

\begin{table}[H]
\caption{Names of the parts in Figure 5.3}
  \begin{center}
    \begin{tabular}{| l|c |}
    \hline % horizontal line demek
    1 & Gas Inlet \\
    \hline
    2 & Plasma Chamber (Pyrex Tube) \\
    \hline
    3 & Extraction Unit \\
    \hline
    4 & Vacuum Gradient  \\
    \hline
    5 & Faraday Cup Unit \\
    \hline
    6 &  Vacuum Line (for the extraction unit) \\
    \hline
    7 &  Vacuum Line (for vacuum gradient) \\
    \hline
    \end{tabular}
  \end{center}
  
 \label{Names of the parts in Figure 5.3}
\end{table}
%buraya sistemin yakından çekilmiş numaralandırılmış resmini koy. numaralar üzerinden anlat.

\section{PLASMA GENERATION BY MICROWAVES}
The microwave ion generation unit of the system comprises of  a waveguide with the length of 375 mm and a pyrex plasma chamber, mounted  through the waveguide horizontally. The circular holes that hold the plasma tube have a diameter of 30 mm in order to mount the pyrex plasma chamber tightly.
 
%5.4
\begin{figure}[H]
 \centering
 \includegraphics[width=0.5\textwidth]{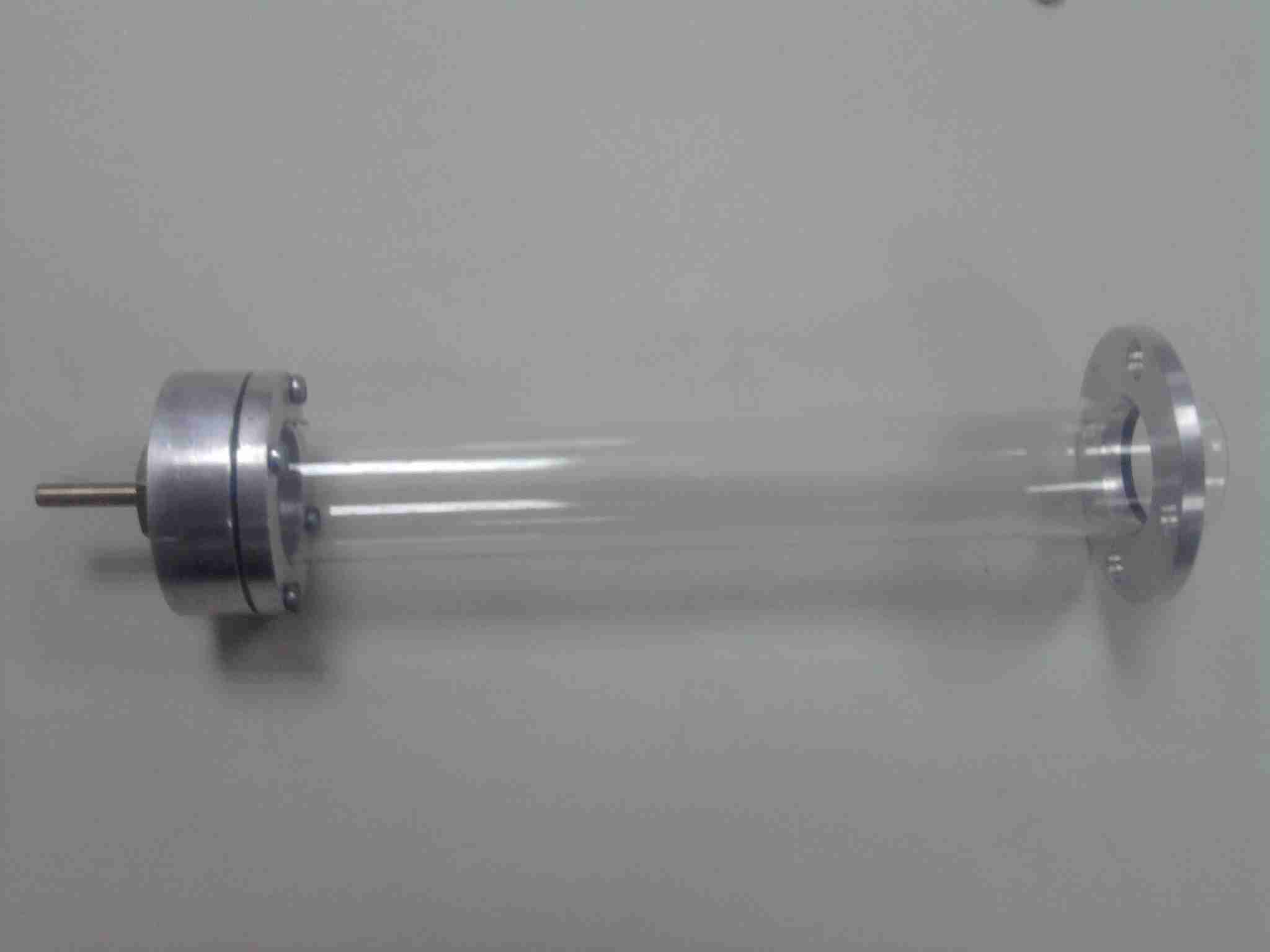}
 \caption{The Pyrex Plasma Chamber; with the gas inlet connector on the left side }
 \label{The Pyrex Plasma Chamber; with the gas inlet connector on the left side }
\end{figure}

\begin{figure}[H]
 \centering
 \includegraphics[width=0.6\textwidth]{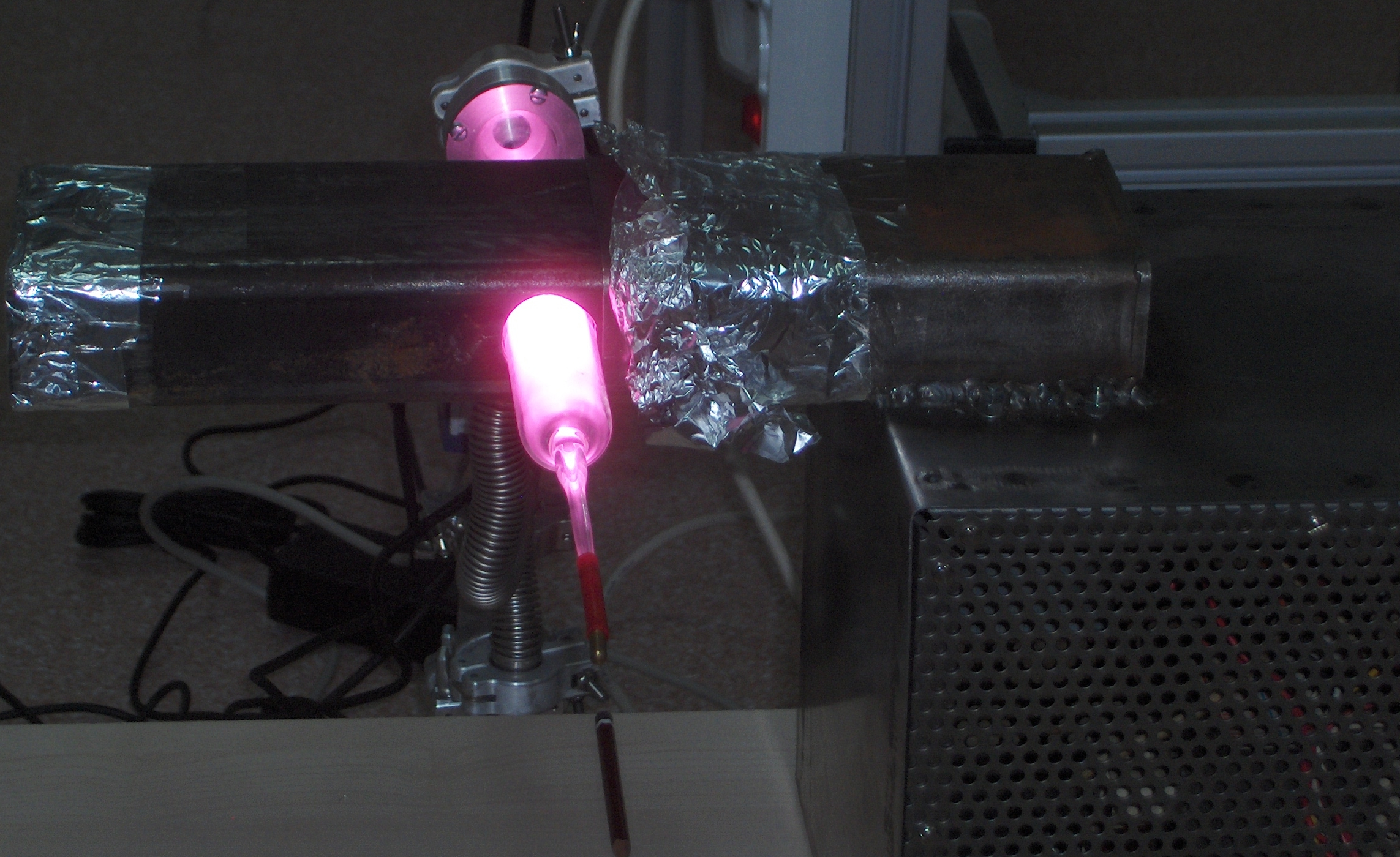}
 \caption{Plasma generation in oder to test the microwave unit}
 \label{Plasma generation in oder to test the microwave unit}
\end{figure}

The ends of the waveguide are closed and sealed in order not only  to produce standing wave pattern inside the waveguide, but also  to prevent   microwave leakage. The generated microwaves are pumped in near the end of the waveguide via the  microwave antenna of the magnetron. The length of the waveguide is approximately 3 times longer than the wavelength of 2.45 GHz microwave. The plasma is generated between the nodes of the standing waves.    In Figure \ref{Plasma generation in oder to test the microwave unit}, the  photograph of the microwave induced plasma system is shown while the microwave induced plasma is forming through the pyrex plasma tube.

%5.5

\begin{figure}[H]
 \centering
 \includegraphics[width=1\textwidth]{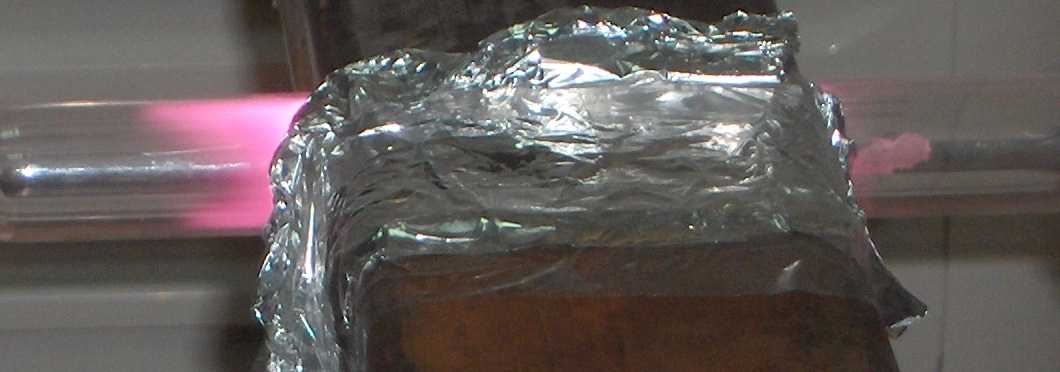}
 \caption{Dispersion zone}
 \label{Dispersion zone}
\end{figure}

 The plasma characterization is done by using a double Langmuir probe. The data of the current is measured by Keithley-2400 Sourcemeter and is collected by Labview-9 via GPIB (General Purpose Interface Bus) interface. The double Langmuir probe is inserted into the plasma at a horizontal position through the cylindrical symmetry axis of the pyrex chamber in order to find out  the plasma parameters, such as the electron density, $n_{e}$, and the electron temperature, $T_{e}$. The data sets are taken at the dispersion zone of the plasma. The dispersion zone is shown in Figure \ref{Dispersion zone}, where the plasma disperses to the ends of the chamber when it pervades from borders where the waveguide encompasses the chamber.

%5.6

\begin{figure}[H]
 \centering
 \includegraphics[width=1\textwidth]{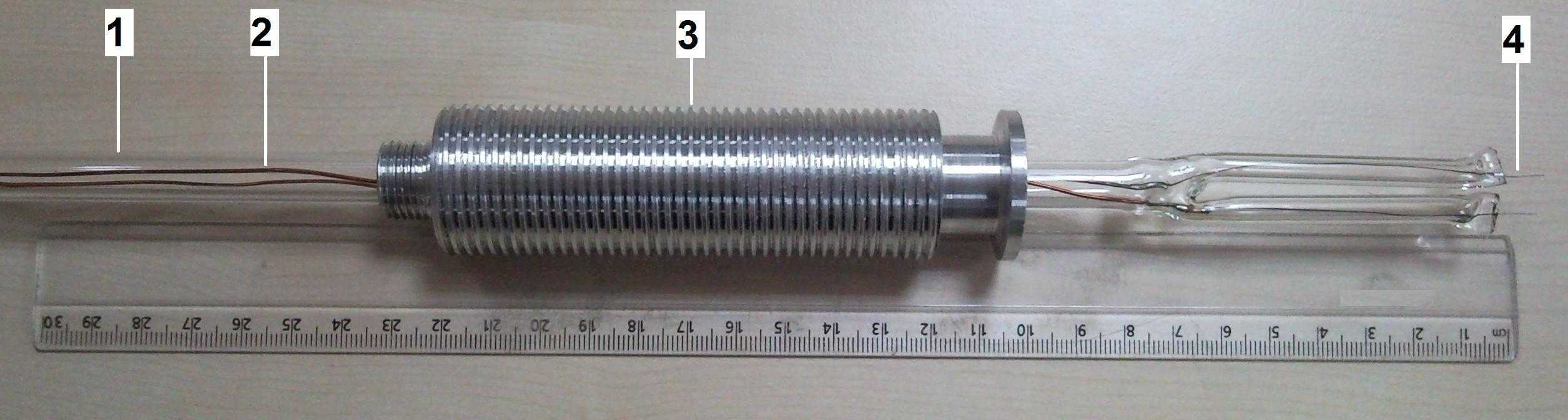}
 \caption{The double Langmuir probe}
 \label{The double Langmuir probe}
\end{figure}

% burdan sonra double probe şelikllerini 2 (belki  3 ) tane ve teorisi ko penning makalesine atıf yparak ve plasma diagnostic kitabına  probe fomullerini anlat. altta kalan fig isimlrini yeni adlandırmaya göre anlat 
In order to measure the plasma parameters, the double Langmuir probe,  used in this system,  is shown in Figure \ref{The double Langmuir probe}. The double Langmuir probe was prepared especially  for this system. The probe has a pyrex probe tube, prepared as  the body (1) of the probe. The ends of the body are sealed. Two copper wire lines (2) cross through the pyrex body and are connected the tungsten wire lines which end at the tips of the probe. The connection  part (3) is  made of aluminum and it connects the probe to the plasma chamber (tube). There is a O-ring, got clamped inside the connection part. This clamped O-ring allows to adjust the location of the probe tips in the plasma chamber by moving through the axis of the body (1) and also prevents gas leakage in between the body(1)  and  the connection part (3). The tips (4) are made of tungsten.

%5.7

\begin{figure}[H]
 \centering
 \includegraphics[width=0.7\textwidth]{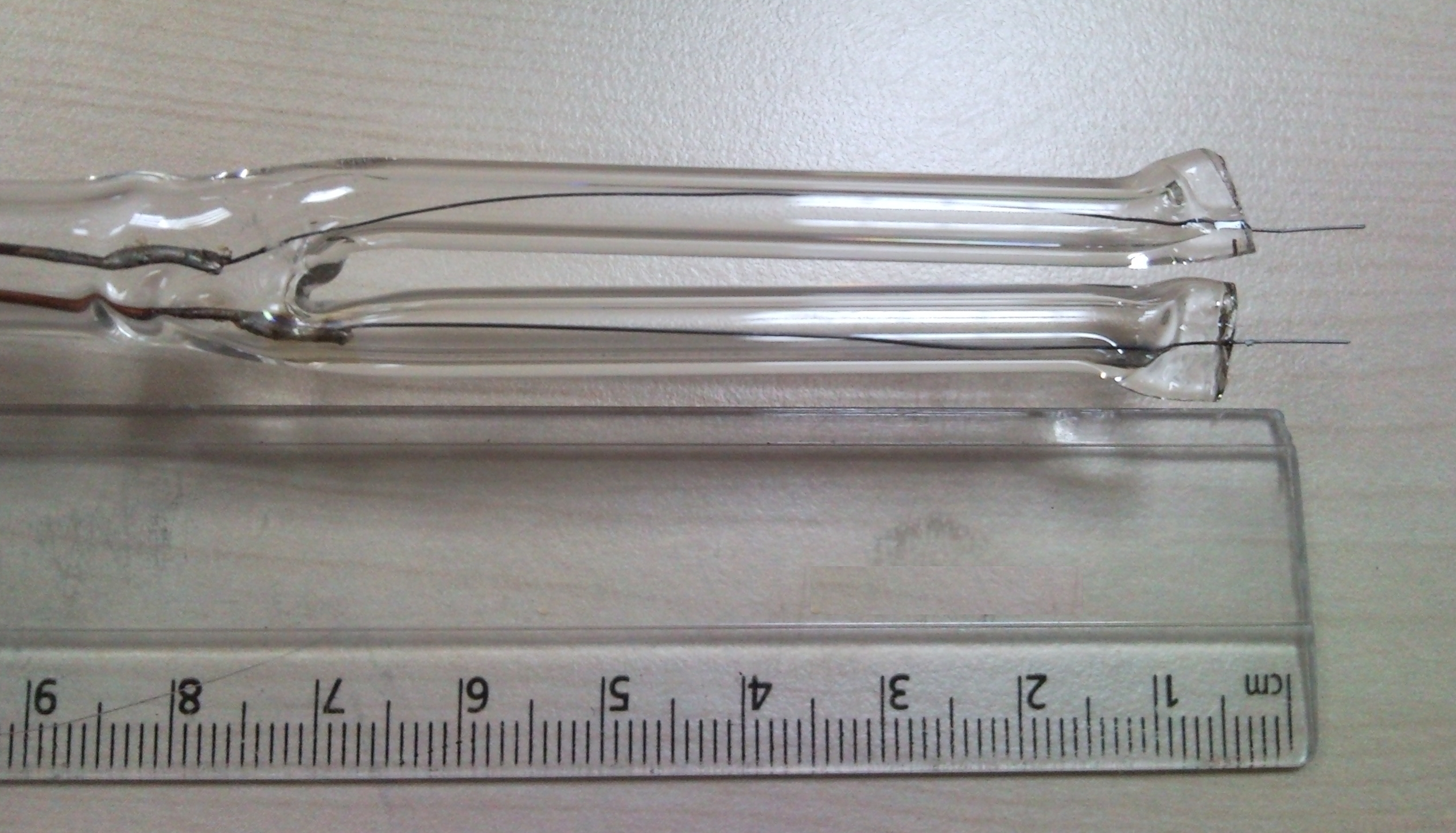}
 \caption{The tips of the probe}
 \label{The tips of the probe}
\end{figure}

There is a closer view of the double Langmuir probe in Figure \ref{The tips of the probe}. The two tips are separated from each other by two pyrex tube branches and the tungsten wires are soldered to the copper wires at the joining point of the two branches. The length of these branches are 6.5 cm long approximately. The diameter of the each probe tip  is 0.25 mm  and the length of the tips is 10 mm. The total surface area of the each probe is $7.9 mm^{2}$.

After the tips of the probe are immersed  into  the plasma, a voltage source is connected to the electrodes of the tips.  The voltage is increased gradually from -70 V to +70 V. The current  and the voltage difference between the two tips are measured and saved at the computer.

%5.8

\begin{figure}[H]
 \centering
 \subfigure[The  plasma is off]{\includegraphics[width=0.4\textwidth]{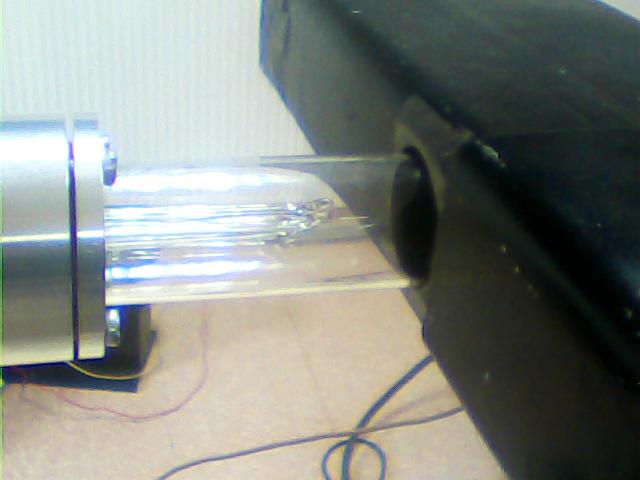}}
 \qquad
 \subfigure[The plasma is on]{\includegraphics[width=0.4\textwidth]{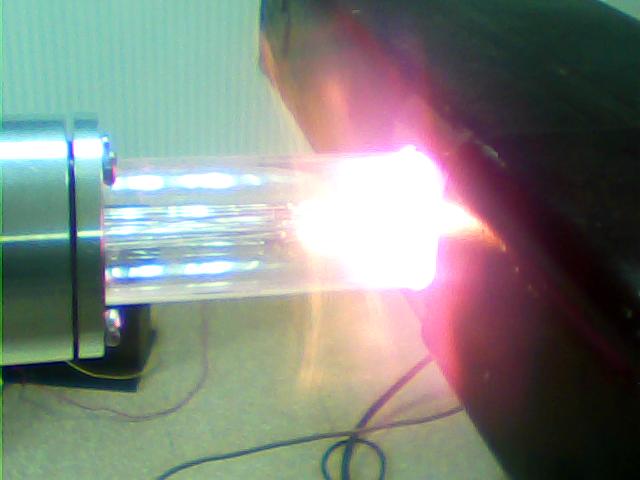}}
 \qquad
 \subfigure[whole view]{\includegraphics[width=0.5\textwidth]{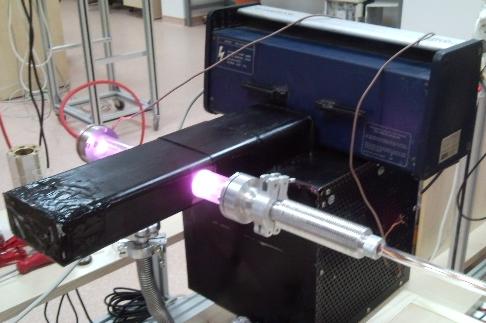}}
 \qquad
 \caption{The location of the probe tips}
 \label{The location of the probe tips}
\end{figure}

%bur araya kadar eklenti.
The tips of the probe are localized at 50 mm away from the center of the waveguide's symmetry axis. The localization of the tips of the probe are shown in Figure \ref{The location of the probe tips}(a) and in Figure \ref{The location of the probe tips}(b). In Figure \ref{The location of the probe tips}(c), the double Langmuir probe is mounted and operated and it can been seen as a silver  color cylindrical metal on the right-hand side of the plasma chamber. In order to prevent the noise  stemming from the  microwave with the frequency of 2.45 GHz, the tips of the probe  are located out side of  the waveguide. If the data are taken at the position  mentioned above, the comparison can be made consistently while the plasma parameters are investigated at the midpoint of the waveguide. Moreover, a filtration process is applied while the plasma is forming. Low pass filter is used at the frequency of 0.1 Hz. 

The values of the electron number density $n_{e}$ and the electron temperature, $T_{e}$ at the midpoint of the waveguide   are expected to be more than the values, measured at the dispersion zone, which is the outside of the waveguide.

%5.9

The current versus voltage graph of the microwave induced plasma, at the pressure of 1.5 mbar (1.13 torr), is shown in Figure \ref{The Current vs Voltage Graph of 1,5 mbar Microwave Hydrogen Plasma at the dispersion zone 1}. The plasma  electron temperature,  $T_{e}$, and the electron density, $n_{e}$,   are calculated by using the following equations \cite{penning}:

\begin{equation}
 \frac{dI_{prob}}{dV_{prob}} \bigg|_{I_{prob}=0}=\frac{1}{2}\frac{eI_{+}}{kT_{e}} \quad,
\end{equation}
\begin{equation}
 T_{e}=\frac{1}{2}\frac{e}{k}\Delta V_{d} \quad,
\end{equation}

where  $A$ is  the area of the individual tip surface, $V_{prob}$ is the voltage difference between the tips of the double probe, $I_{prob}$ is the probe current, $I_{+}$ is the saturation current, $k$ is the Boltzmann constant, and  $m_{+}$ is the mass of the hydrogen ion $H^{+}$.  

%5.10
\begin{figure}[H]
 \centering
 \includegraphics[width=0.8\textwidth]{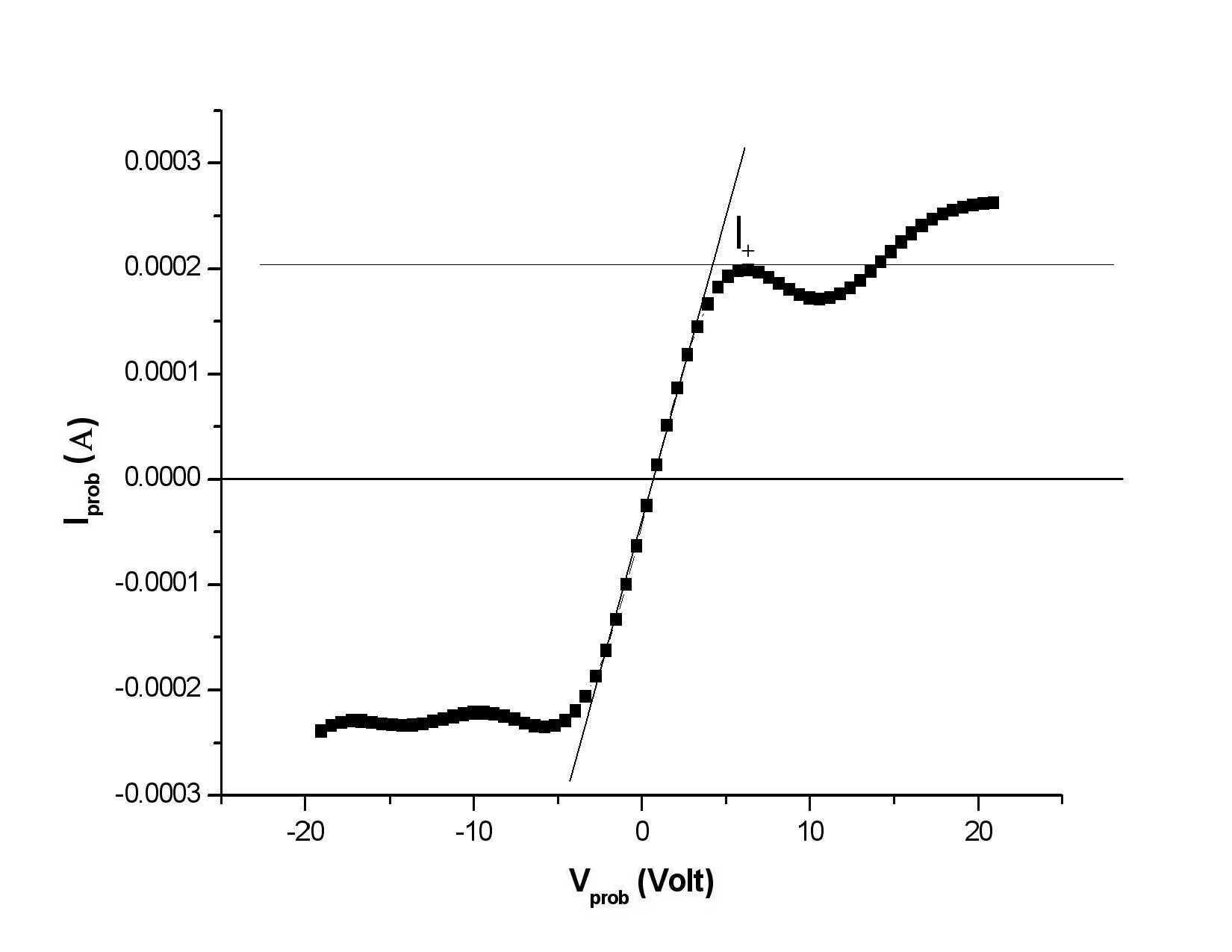}
 \caption{The Current vs Voltage Graph of \textit{1,5 mbar Microwave Hydrogen Plasma at the dispersion zone 1}}
 \label{The Current vs Voltage Graph of 1,5 mbar Microwave Hydrogen Plasma at the dispersion zone 1}
\end{figure}

By using the graph, above, the probe current $I_{prob}$ is obtained as $\sim 250 \mu A$ and the ion saturation current, $I_{+}$, is determined as $\sim 200 \mu A$. In order to determine the slope of the linear part of the curve at $I_{prob}=0$, the value of the  probe current, at $ 250 \mu A $, and the value of the probe voltage,  $V_{prob}$, at $4.38 V$ can be used. By substituting the values of the probe current $I_{prob}=\sim 250 \mu A$ and the corresponding voltage,  $V_{prob}=4.38 V$, into Equation 5.1, the expression can be written as 

\begin{equation}
 \frac{250\times10^{-6} }{4.38 }=\frac{1}{2} \frac{e}{ k  T_{e}} \times 200\times10^{-6}
\end{equation}

If the above equation is arranged for $T_{e}$, then $T_{e}$ is found approximately  $ \sim 2eV $  ($ \sim 23200 ^{o} K $).
\begin{equation}
 n_{e}=\frac{I_{+}}{0.61 e A \sqrt{k T_{e}/m_{+}}}
\end{equation}

The ion saturation current, $I_{+}$, can be determined  by drawing a tangential line horizontally to the first crest of the ion current decline at the positive voltage region. This is done for fitting and  determining the tangent hyperbolic characteristic of the double probe's \textit{I vs V curve}. It should not be forgotten that this procedure is approximate determination of the ion saturation current.
If the ion saturation current, $I_{+}$, is $\sim 200 \mu A$ is substituted into Equation 5.4 \cite{penning}, Equation 5.4 becomes

\begin{equation}
 n_{e}= \frac{ 200\times10^{-6} }{0.61 \times 1.6\times10^{-19} \times 7.9\times10^{-6} \sqrt{ 1.38\times10^{-23} \times 23200 / 1.6\times10^{-27}}}
\end{equation}

Finally, the electron density, $n_{e}$, is calculated for the pressure of 1.5 mbar to be  $2\times10^{16} m^{-3}$ or $2\times10^{10} cm^{-3}$.

\begin{figure}[H]
 \centering
 \includegraphics[width=0.8\textwidth]{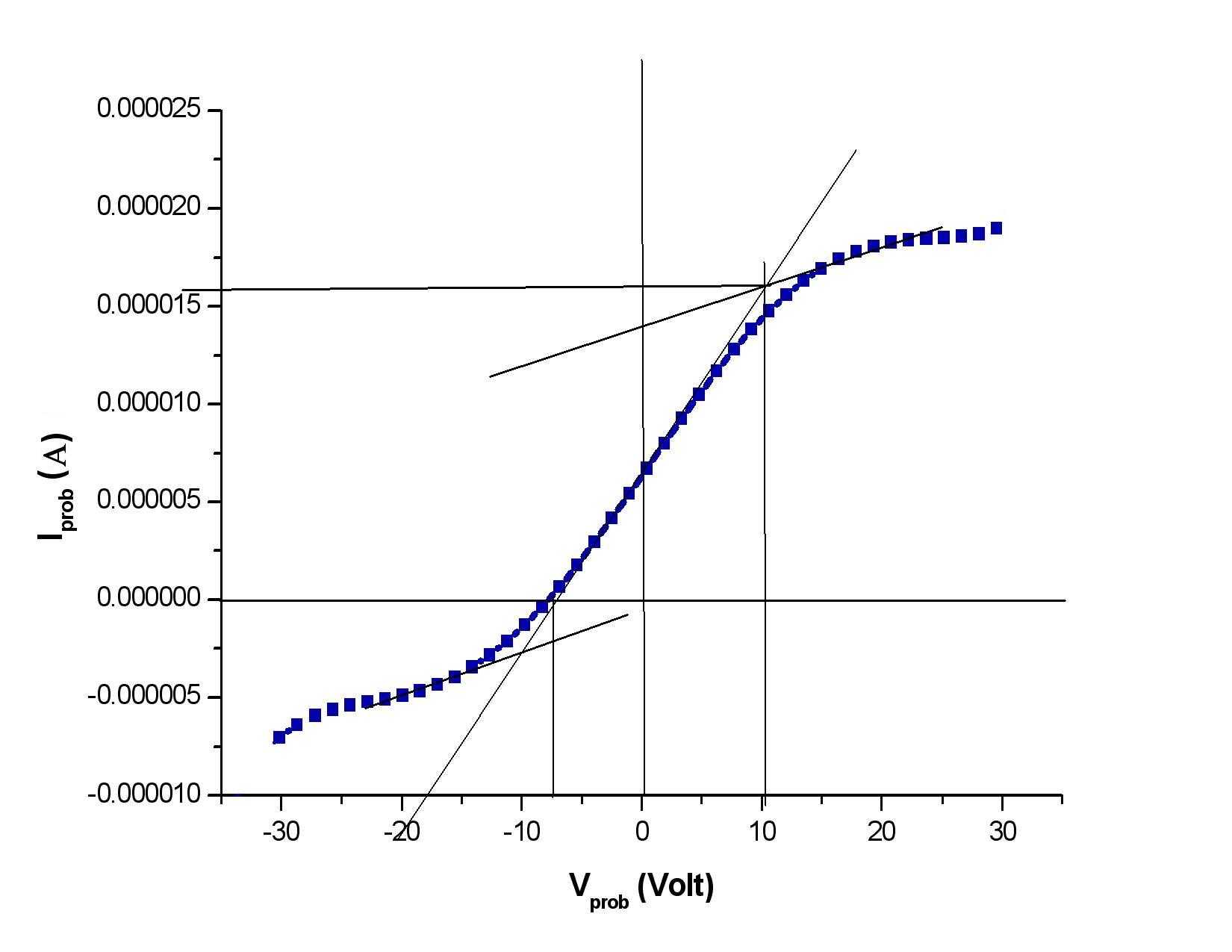}
 \caption{The Current vs Voltage Graph of \textit{1.5 mbar Microwave Hydrogen Plasma at the midpoint of the waveguide} }
 \label{The Current vs Voltage Graph of 1.5 mbar Microwave Hydrogen Plasma at the midpoint of the waveguide}
\end{figure}

By repeating the experiment, the following data were taken and the graph is shown in Figure \ref{The Current vs Voltage Graph of 1.5 mbar Microwave Hydrogen Plasma at the midpoint of the waveguide}, in order to obtain the plasma electron temperature and density at the midpoint of the waveguide.  The pressure was held at  1.5 mbar and again the hydrogen gas was used. By using the graph, in Figure \ref{The Current vs Voltage Graph of 1.5 mbar Microwave Hydrogen Plasma at the midpoint of the waveguide}, and by doing the same calculations, mentioned previously, the electron temperature,  $T_{e}$, is obtained as $\sim 8eV$ ($\sim 92800 ^{o} K$), and the electron density, $n_{e}$, is obtained in  the order of  $\sim 7.5\times10^{8} cm^{-3}$. As it can be seen from the value of the electron number density,  $n_{e}$, it is smaller than the value which is found for the data of the dispersion zone,  is in the order of $10^{10} cm^{-3}$.
 
%5.11

 As it is seen in Figure \ref{The Current vs Voltage Graph of 1.5 mbar Microwave Hydrogen Plasma at the midpoint of the waveguide}, the \textit{I vs V curve} of the probes is not symmetric about the origin. And also the found ion density is smaller then the \textit{dispersion zone 1}. It is not an expected value for the core of the plasma.  
 
\begin{figure}[H]
 \centering
 \includegraphics[width=0.8\textwidth]{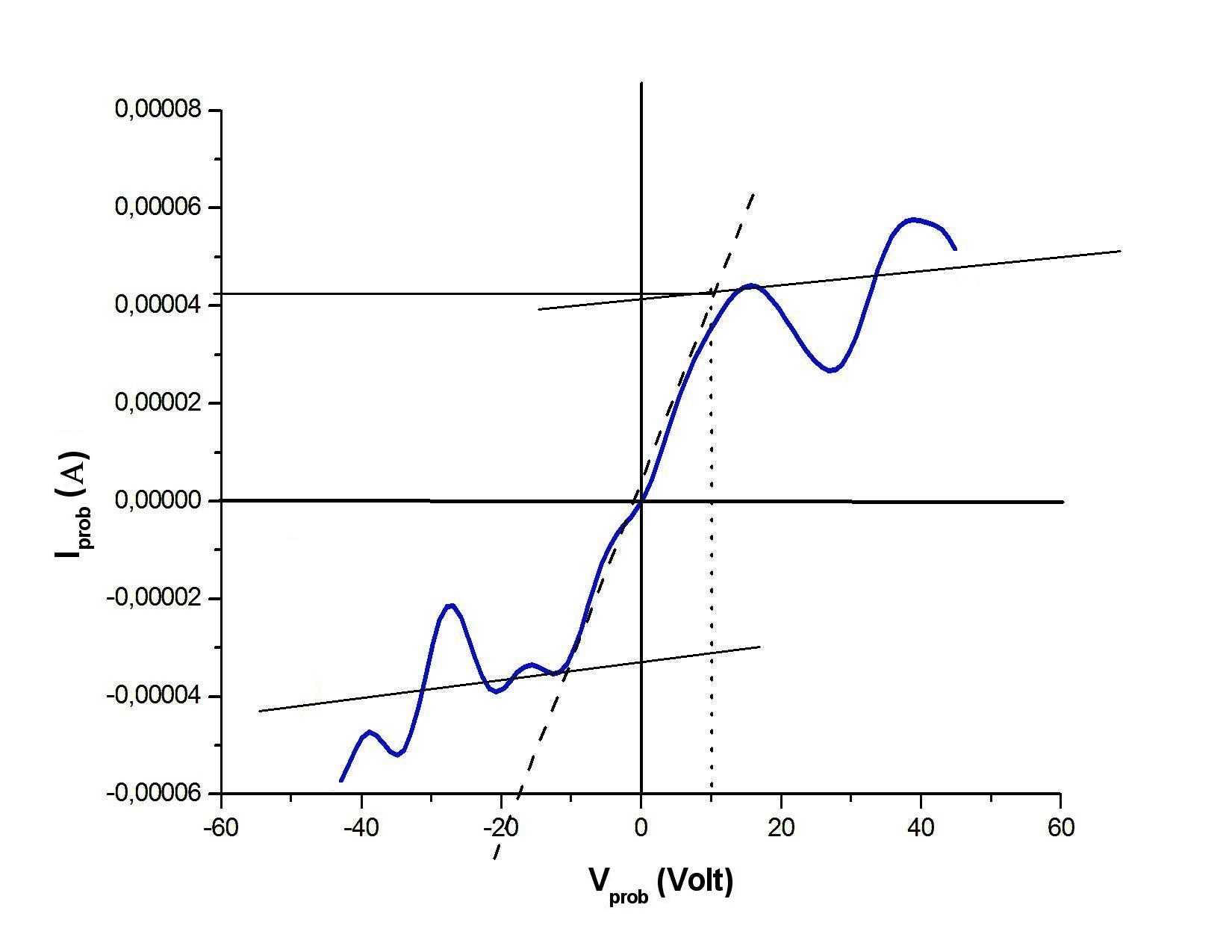}
 \caption{The Current vs Voltage Graph of \textit{1.5 mbar Microwave Hydrogen Plasma at the dispersion zone 2}}
 \label{The Current vs Voltage Graph of 1.5 mbar Microwave Hydrogen Plasma at the dispersion zone 2}
\end{figure}

 The experiment was repeated again in order to obtain the symmetric \textit{I vs V curve} of the plasma  outside of the waveguide. The probe tips were at  a distant of 57 mm from the midpoint of the waveguide. The reason of this is to observe the electron density decrease while moving the probe to the outside of the waveguide. The data were taken for the microwave hydrogen plasma at the pressure of 1.5 mbar.  The  graph of this process is shown in Figure \ref{The Current vs Voltage Graph of 1.5 mbar Microwave Hydrogen Plasma at the dispersion zone 2}.

%5.12

 If the  data analysis  above  is applied, the electron temperature,   $T_{e}$, is obtained as $\sim 5eV$ ($\sim 58000^{o} K$), and the electron density, $n_{e}$, is obtained in the order of  $\sim 10^{9} cm^{-3}$ at the distant, 57 mm far from the midpoint of the waveguide and in the vicinity of the extraction electrodes.  

 As it can be  seen from the result of the calculation, the electron density of the plasma  decreases when the plasma disperses. However, the second \textit{I vs V Graph}  in Figure \ref{The Current vs Voltage Graph of 1.5 mbar Microwave Hydrogen Plasma at the midpoint of the waveguide}  is not as accurate as the first  \textit{I vs V Graph}, in Figure \ref{The Current vs Voltage Graph of 1,5 mbar Microwave Hydrogen Plasma at the dispersion zone 1}. The reason of this is that the second \textit{I vs V Graph} is not symmetric  with respect to the origin and there is a shift on the graph. While the data are been taken out side of the waveguide, at \textit{the dispersion zone 2}, the plasma can be  affected by the environmental impacts. And also at a little far distant, out of the waveguide, the plasma  becomes more instable. Since the microwaves can not provide the sufficient energy to the molecules in order to ionize them at that region. The plasma   starts  flickering  far from the waveguides borders. The waved shapes, at the left and right sides of the curve, can be caused by these instabilities at \textit{the dispersion zone 2}.  The region, where the data are taken accurately is very narrow in this experiment. However, there is some noise, observed in the waveguide  that is not eliminated. Consequently,  for  the data of the dispersion zones, the current readings and the voltage readings are more accurate than the data of the midpoint zone. While taking the data sets of the experiment, the noise factors are discarded by low pass filters via  the computer. The filter tools are used to get rid of the noise factors in the computer program. However, these filtering tools are not electronic tools, they are only the computational tools that reduce some of the unwanted signal while the \textit{I vs V Graphs} are being drawn and they are not enough to get rid of the noise and the other effects which occurred at the midpoint of the waveguide. The other effects  are discussed at the conclusion part of the thesis.  While doing the experiment, the data, which were obtained for the 1.5 mbar microwave plasma at the dispersion zones, have been used in the calculations and comments. 

\newpage
\section{EXTRACTION UNIT}
\begin{figure}[H]
 \centering
 \includegraphics[width=1\textwidth]{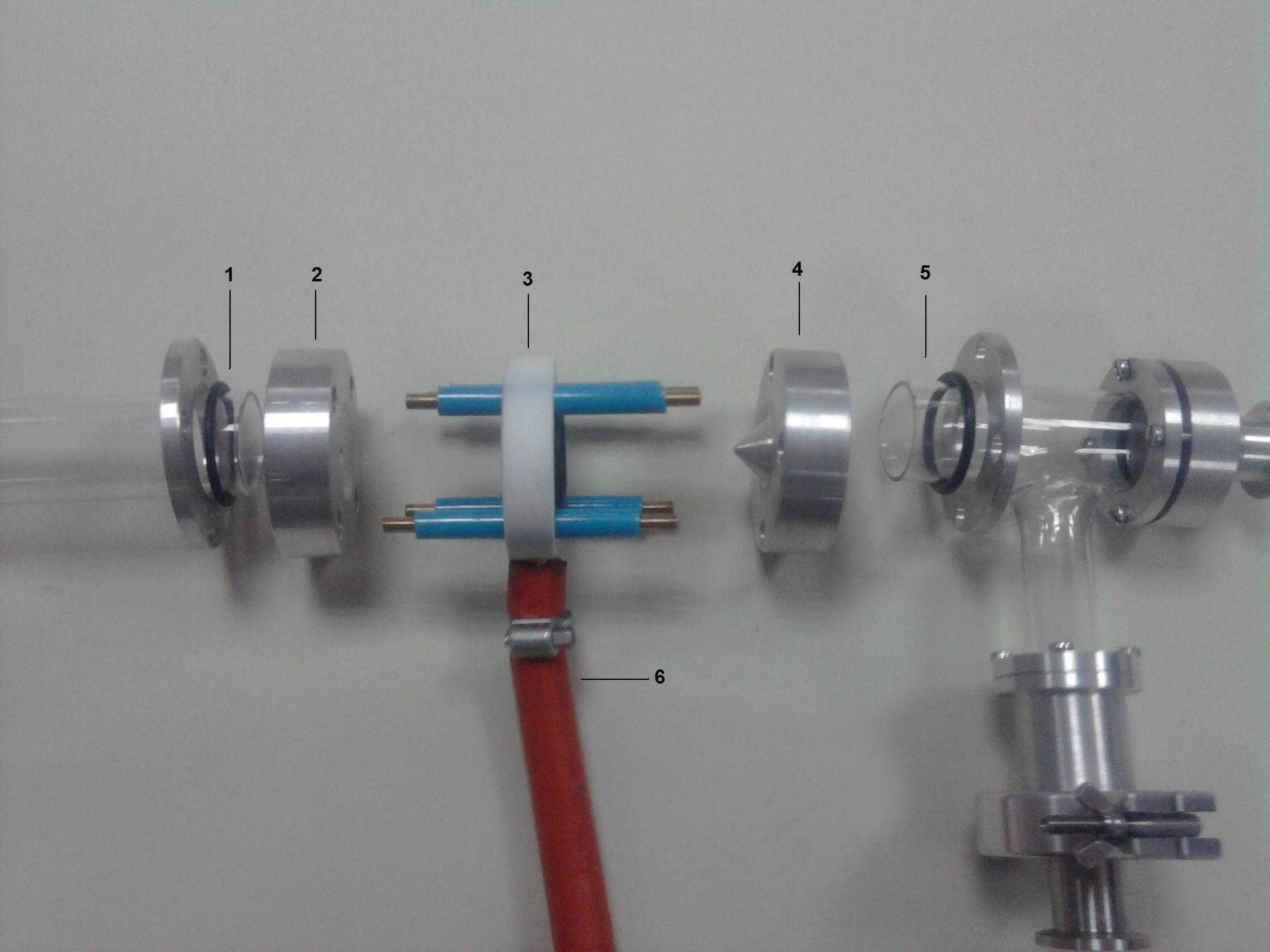}
 \caption{Disassembly of the extraction unit}
 \label{Dissembly of the Extraction Unit}
\end{figure}

 In this experiment, the diode extraction system is used with the single aperture.  There is a disassembled view of the extraction unit in Figure \ref{Dissembly of the Extraction Unit} and the parts are named as in Table \ref{Explanation of Figure 5.13}. 

%5.13
  
\begin{table}[H]
\caption{Explanation of Figure 5.13}
  \begin{center}
    \begin{tabular}{| l|c |}
    \hline % horizontal line demek
    1 & Plasma Chamber Connection \\
    \hline
    2 & First Single Aperture \\
    \hline
    3 & Teflon Separator \\
    \hline
    4 & Conical Single Aperture   \\
    \hline
    5 & Vacuum Gradient Connection \\
    \hline
    6 &  Vacuum Line \\
    \hline
    \end{tabular}
  \end{center}
  
 \label{Explanation of Figure 5.13}
\end{table}
 
 The plasma chamber connection is made of aluminum  in the shape of a disk which joins the first single aperture with four bolts and an O-ring is placed in between the plasma chamber connection and the first single aperture. All the o-rings are lubricated with low pressure lubricant.

\begin{figure}[H]
 \centering
 
 \subfigure[Plasma Side]{\includegraphics[width=0.3\textwidth]{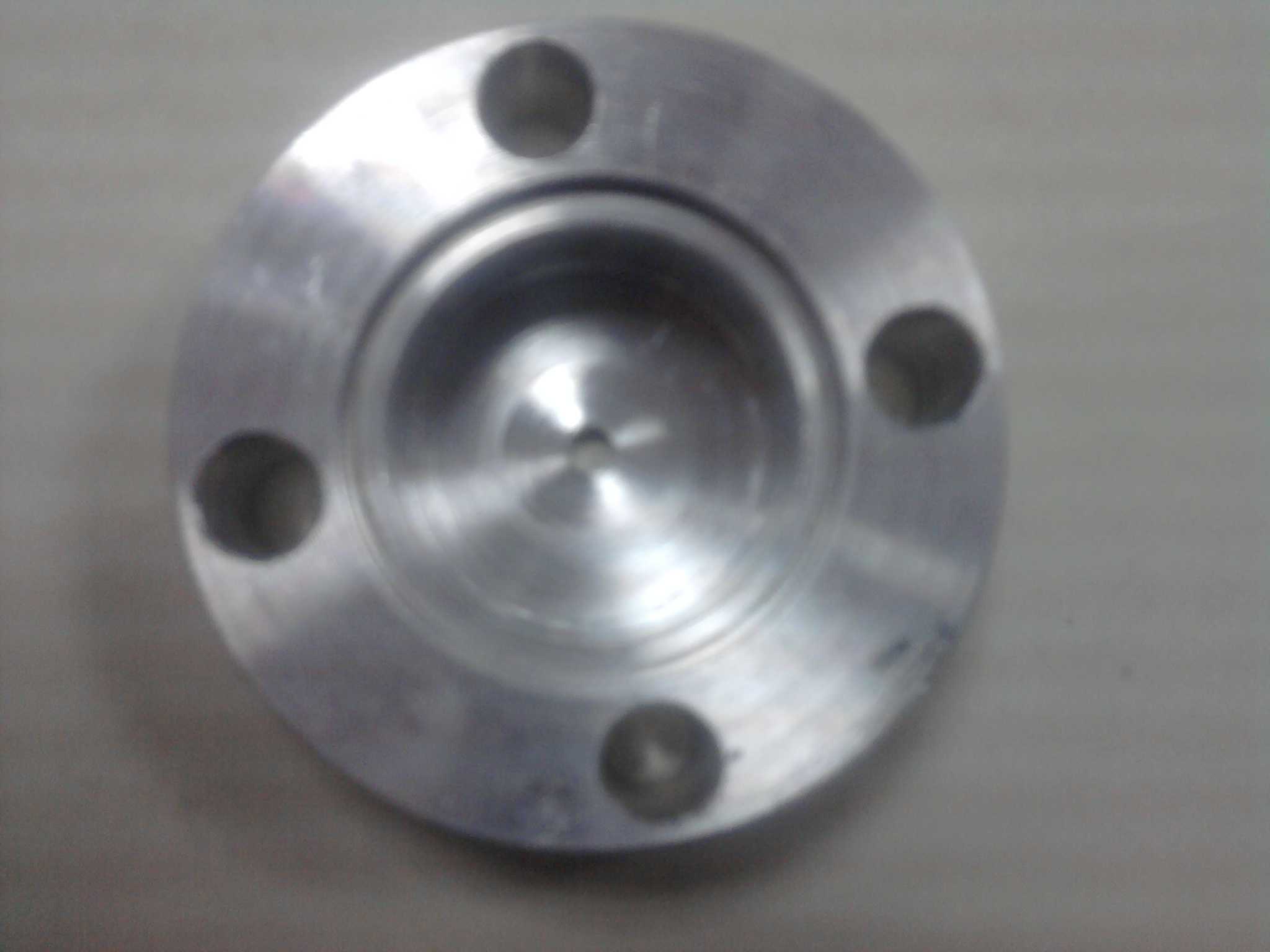}}
 \qquad
 \subfigure[Teflon Separator Side]{\includegraphics[width=0.3\textwidth]{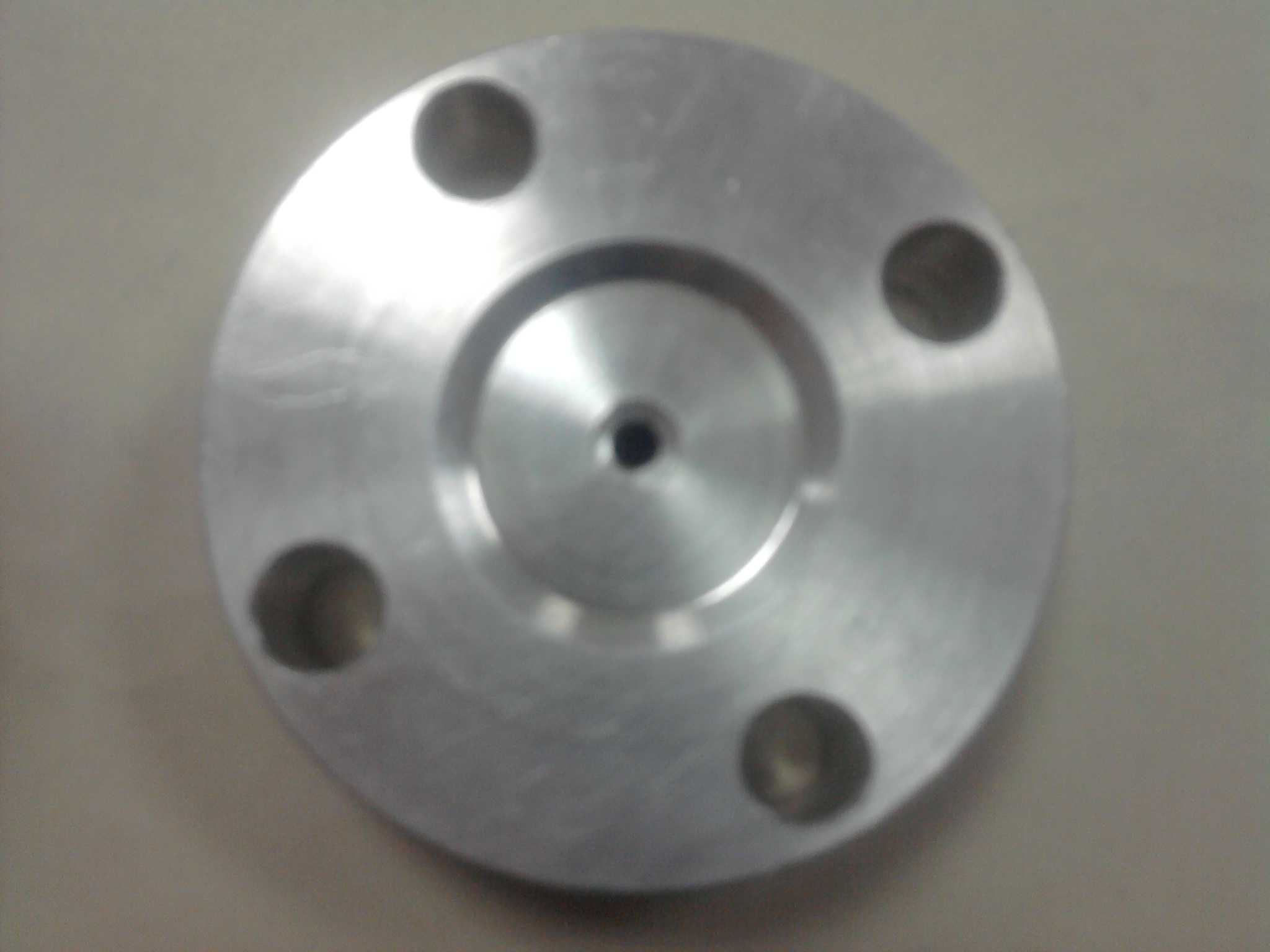}}
 \qquad

 \caption{The First Single Aperture}
 \label{The First Single Aperture}
\end{figure}

 In Figure \ref{The First Single Aperture}(a), the plasma side of the first single aperture is shown. This side faces the plasma chamber and the plasma chamber is mounted in this side. In Figure \ref{The First Single Aperture}(b), the teflon separator side is shown. This side faces the teflon separator and presses the teflon separator with four long bolts which connect the first single aperture, the teflon separator and the conical single aperture together. The side, facing the teflon separator, is smooth and flat to provide a proper vacuum. The diameter of the aperture is 4 mm and the thickness of the aperture has a width of 1 mm, approximately.

%5.14

\begin{figure}[H]
 \centering
 
 \subfigure[Profile]{\includegraphics[width=0.3\textwidth]{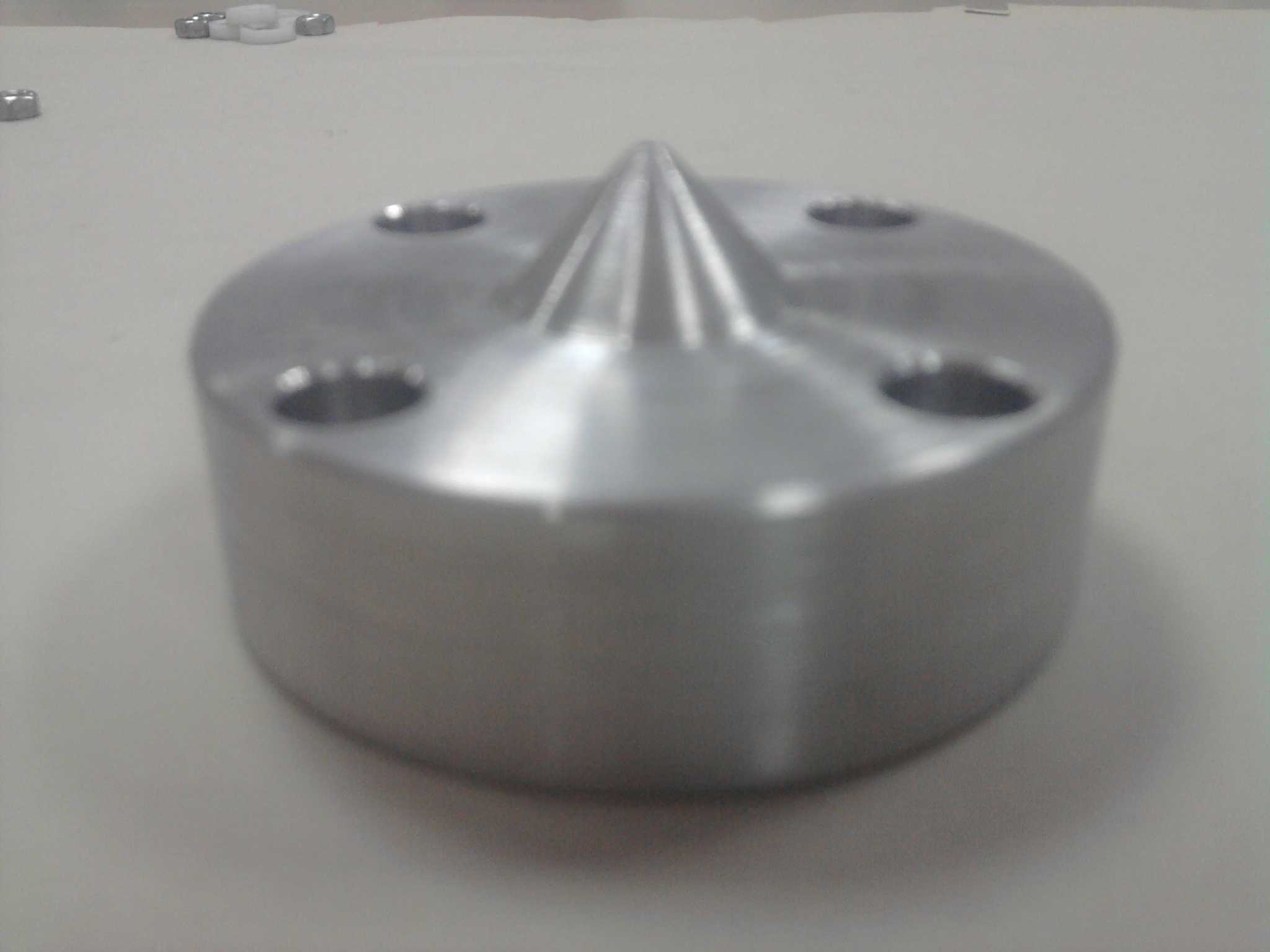}}
 \qquad
 \subfigure[Teflon Separator Side]{\includegraphics[width=0.3\textwidth]{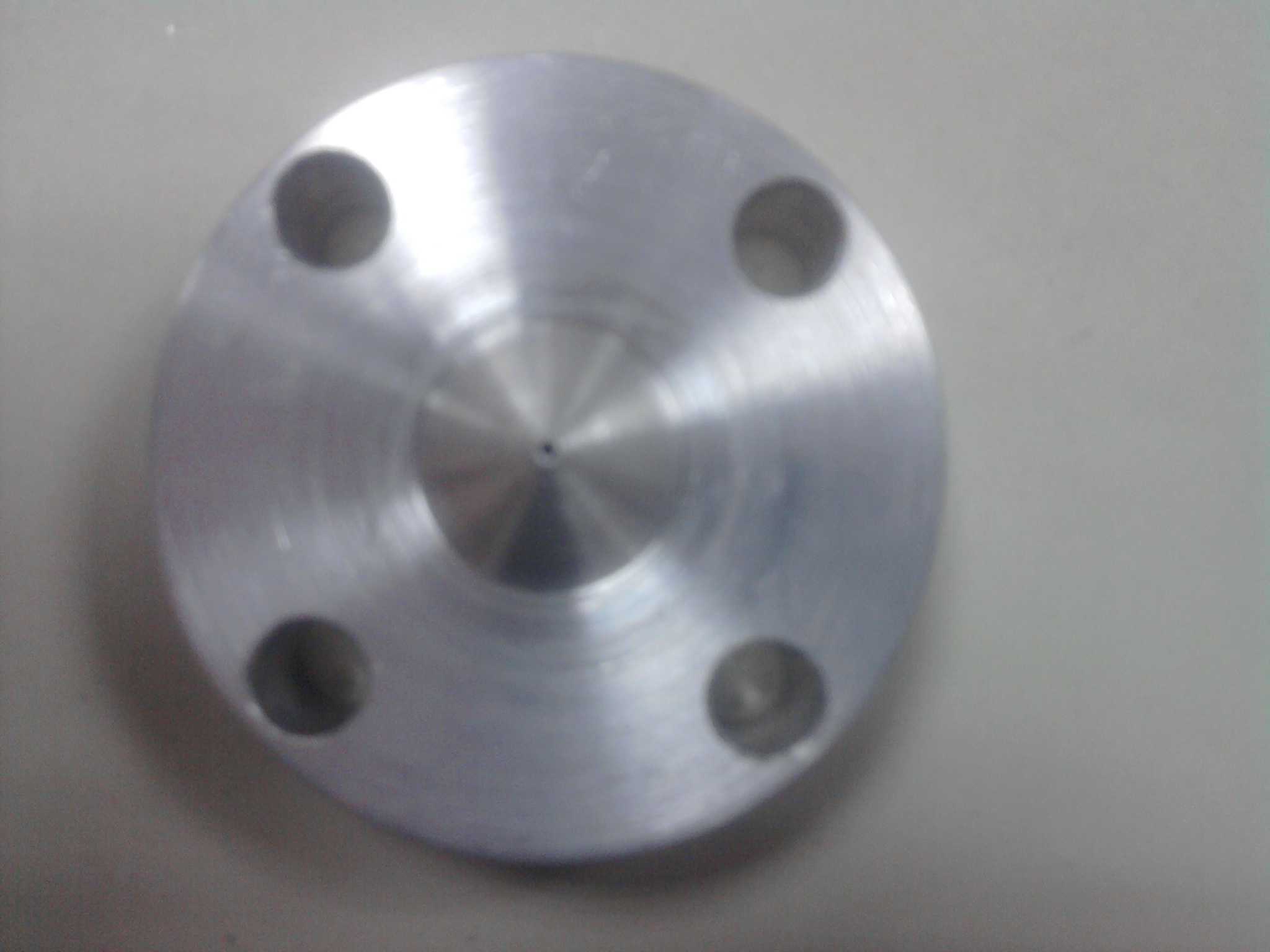}}
 \qquad
 \subfigure[Vacuum Gradient Side]{\includegraphics[width=0.3\textwidth]{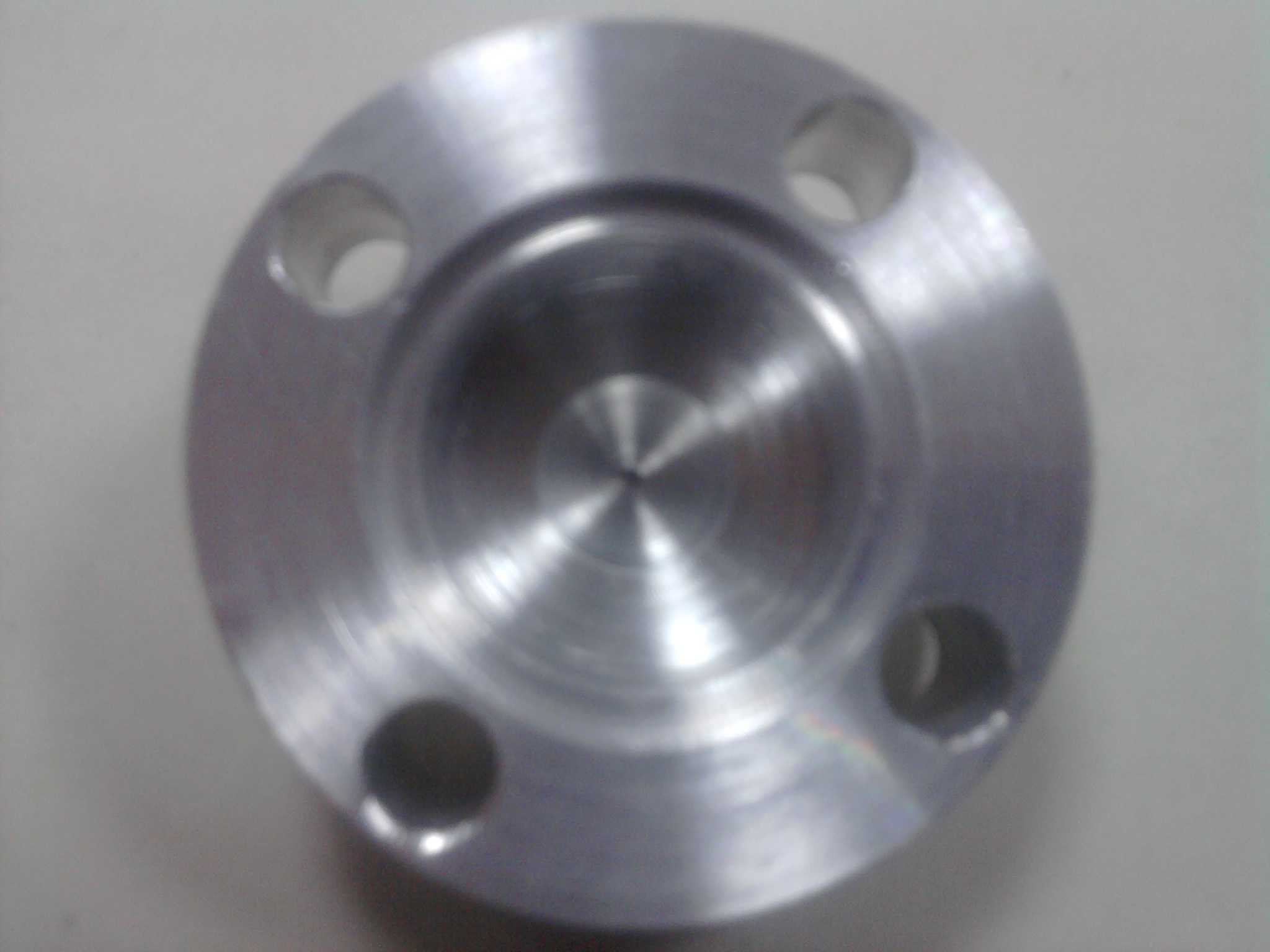}}
 \qquad

 \caption{The Conical Single Aperture}
 \label{The Conical Single Aperture}
\end{figure}

 The  conical single aperture is shown in Figure \ref{The Conical Single Aperture}. The conical side faces the teflon separator and it has an aperture, with the diameter of 2 mm, at the tip. The reason of the conical shape is to produce curved electric field lines in order to guide the accelerated ions to the aperture. In Figure \ref{The Conical Single Aperture}(b), the teflon separator side of the conical aperture is shown. This side  has a smooth and flat surface, like the teflon separator side of the first single aperture, in order to make perfect contact with the teflon separator. The other side of the conical single aperture has a cavity (in Figure \ref{The Conical Single Aperture}(c)), like the first single aperture has also, in order to join the vacuum gradient part. And  this side is mounted to the vacuum gradient part with an O-ring and a aluminum connection disk. 
 Both the first single aperture and the conical single aperture are used as the electrodes in order to apply the extraction voltage.

\begin{figure}[H]
 \centering
 \includegraphics[width=0.5\textwidth]{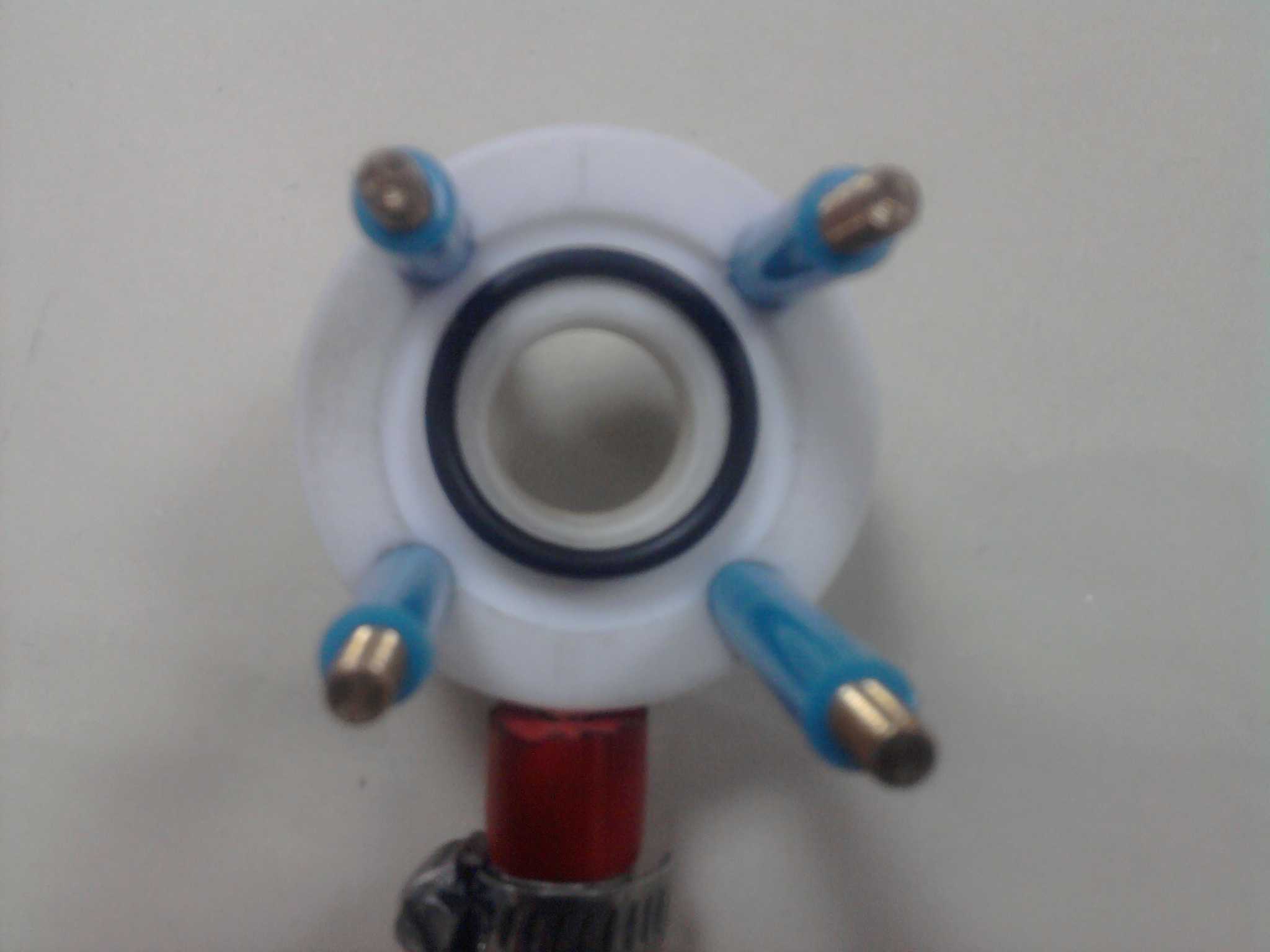}
 \caption{The Teflon Separator}
 \label{The Teflon Seperator}
\end{figure}

 In Figure \ref{The Teflon Seperator}, one side of the teflon separator is shown. The teflon separator is used for an isolator in between the two extraction electrodes, which are also the apertures of the extraction unit. Besides of being used as an isolator, it is very durable for the vacuum and  is convenient to be processed and shaped. The teflon separator has O-rings and cavities on its both sides. The O-rings are placed in the coves and they make perfect connections with the first single aperture and the conical single aperture. The teflon separator also has a vacuum output individually at the bottom of it. The vacuum output is also shown as a red vacuum pipe in Figure \ref{The Teflon Seperator}. The blue plastic pipes are used to isolate the bolts and nuts in order to prevent any short circuits between the first single aperture and the conical single aperture.

%5.15

%5.16

\begin{figure}[H]
 \centering
 \includegraphics[width=0.5\textwidth]{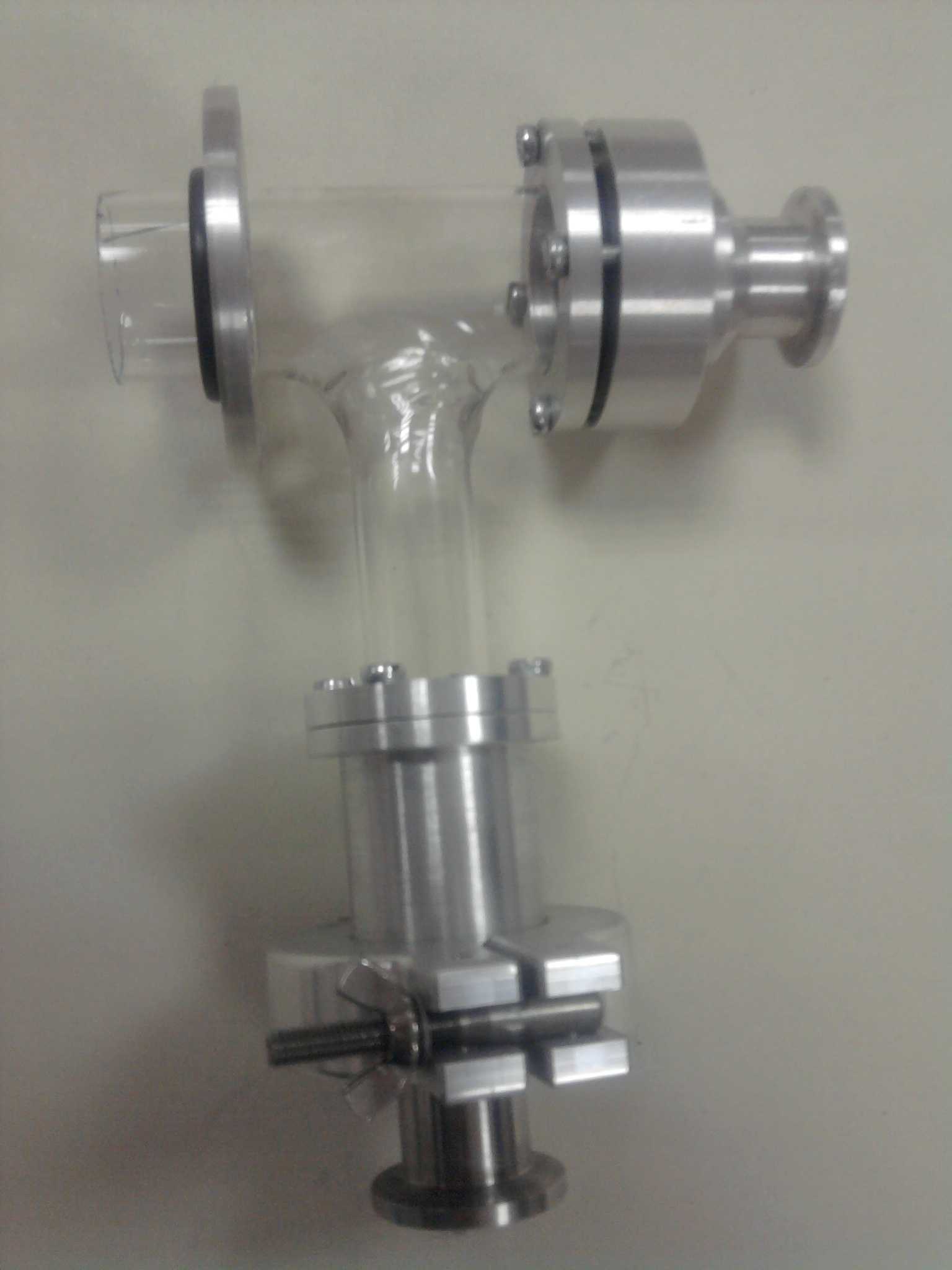}
 \caption{The Vacuum Gradient Part}
 \label{The Vacuum Gradient Part}
\end{figure}

 The vacuum gradient part is shown in Figure \ref{The Vacuum Gradient Part} with its connection disks and a vacuum pump connector. The left side of the vacuum gradient part is connected to the conical single aperture with an O-ring and a connection disk. The right side of it is connected to the Faraday cup in order to measure the ion beams. In addition to them, the connection of the underside joins with the high vacuum pump in order to produce very low pressure gradient. While the pressure, inside the teflon separator, is in between $10^{-1}$ and $10^{-2}$ mbars, the inside of the vacuum gradient part can be at the pressure of the order of $10^{-3}$ mbars.

 Before starting the experiment, the whole system is cleaned in order to use hydrogen gas in the experiment. After that, the whole system pressure is brought to a desired pressure and then the vacuum gradient side is brought to lower pressures to elongate the mean free path of the ions. So they can travel more distance without collisions.  If the inside of the teflon separator is held at a constant pressure, the vacuum gradient will be sustained through the whole system, from the gas inlet to the target.

%5.17

%neden vacuum gardient yarattın esimated paschen curve olayından bahset.
 The vacuum gradient part plays a very crucial role in this experiment. If the pressure is not low enough inside the vacuum gradient part, which takes place in between the conical single aperture and the Faraday cup,  a DC discharge can be induced  between the conical single aperture electrode and the metal flange of the Faraday cup. It is  important to prevent this DC discharge for measuring the accurate data of ion current, produced by the microwave induced plasma and its corresponding extraction unit. The pressure, in between the extraction electrodes, changes  from the order of $10^{-1}$ mbars to the order of $10^{-2}$ mbars (from the order of $10^{-2}$ torr to the order of $10^{-3}$ torr). The estimated Paschen curve of the hydrogen gas, in Figure 2.4, is used for determining the critical value of the DC breakdown. If the highest pressure value, $10^{-2}$ torr, is taken into account, the value of $pd$ becomes $ 5\times10^{-3}$ torr-cm for the minimum separation of 5 mm, between the extraction electrodes. If the corresponding breakdown voltage is found from Figure 2.4, it is seen that the value of the breakdown voltage  exceeds the value of  10 kV. In the designed system, merely 3 kV is applied  between the extraction electrode and the Faraday cup. Therefore, the occurrence of the DC discharge plasma is prevented.

\newpage
\section{MEASUREMENTS OF ION CURRENT}
While measuring the ion current, a Faraday cup is used. A Faraday cup is a device which measures the current of charged particles. The basic theory of the Faraday cup is that a cup shaped metal faces the incoming charged particles.  The impinging  charged particles charge up the Faraday cup and it gains  a net charge. The Faraday cup discharges this small amount of the ion accumulation by grounding. If this small amount of current is measured by an electrometer then the current of the ions, entering  the Faraday cup, is determined. The number of ions per second, entering the Faraday cup, $N$ is given by,
\begin{equation}
 N=It/e \quad,
\end{equation}
where $I$ is the measured current (in  Amperes), $t$ is the time interval(in seconds) when the measurement is done, and $e$ is the electronic charge (in Coulombs). The above equation can be assumed as a good approximation for the ions, having the ionization state of one. 

%5.18
\begin{figure}[H]
 \centering
 \includegraphics[width=0.6\textwidth]{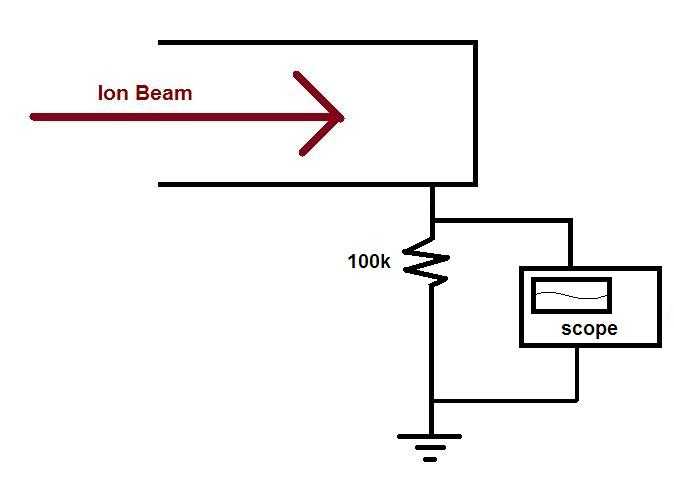}
 \caption{The Faraday Cup Scheme}
 \label{The Faraday Cup Scheme}
\end{figure}

If a resistor is connected between the Faraday cup and the ground in series, the voltage drop can be measured over this resistor and the amount of current can be calculated in order to neutralize the incoming beam, impinging into the Faraday cup. An electrometer can also be connected in series instead of the resistor. The Diagram is depicted in Figure \ref{The Faraday Cup Scheme}. Figure \ref{The Faraday Cup Scheme} also shows the principal circuit diagram of the measurement unit of the microwave ion source. In this system, there is a resistor with a resistance of  $100$ $ k\Omega $ in series and the voltage is read on this resistor by a scope.

% faraday cubın özelliklerinden bahset. yaptığın deneyde nasıl ölçümler aldın ve simulasyon karşılaştırmasını göster.resimlerde anlat ölçüm sonuçlarını.

\begin{figure}[H]
 \centering
 \includegraphics[width=0.8\textwidth]{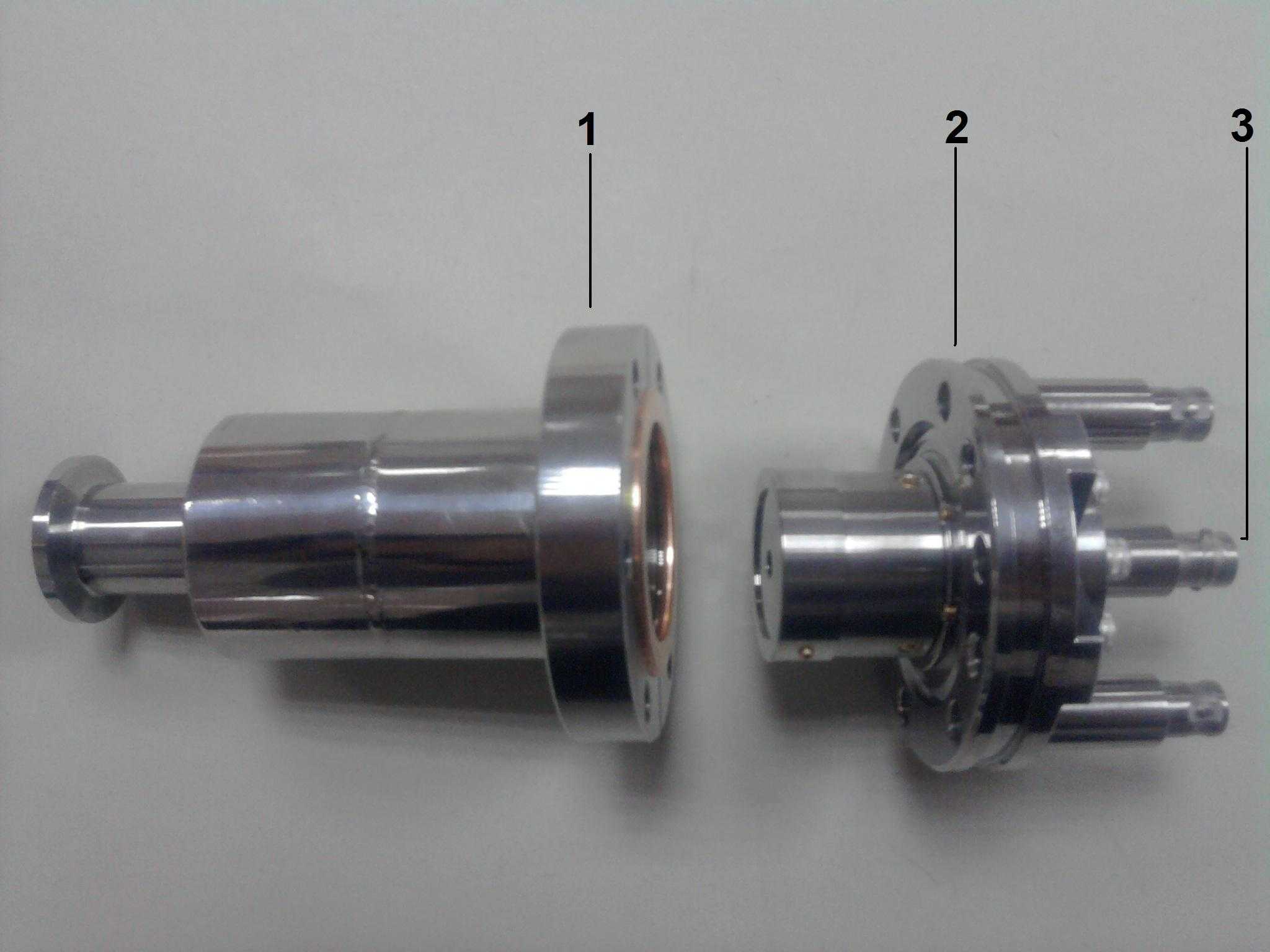}
 \caption{The Faraday Cup (side view)}
 \label{The Faraday Cup (side view)}
\end{figure}

\begin{figure}[H]
 \centering
 \includegraphics[width=0.9\textwidth]{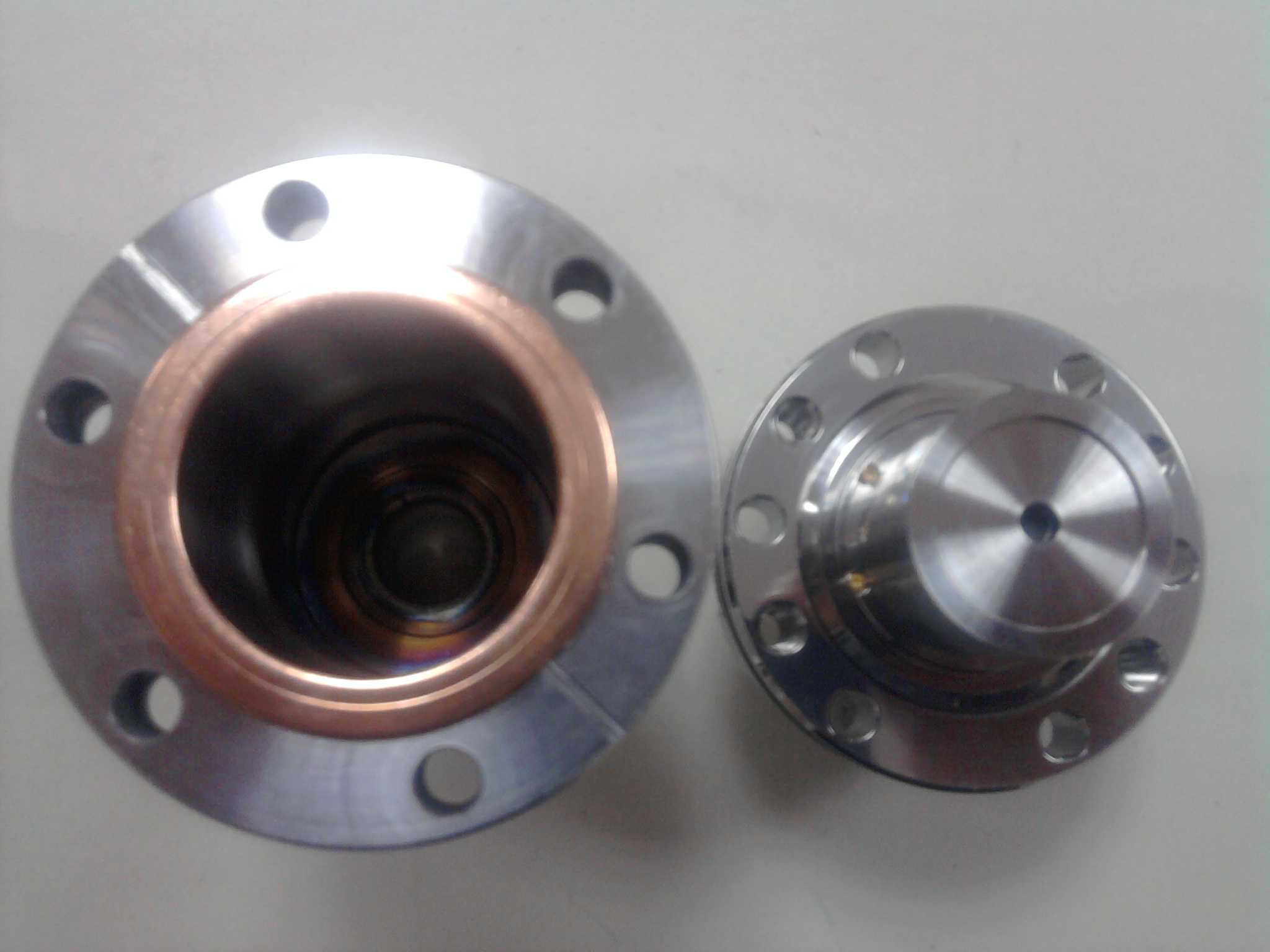}
 \caption{The Faraday Cup (inside view)}
 \label{The Faraday Cup (inside view)}
\end{figure}

For the measurements of this microwave ion source, the Kimball Physics model FC-73A Faraday cup is used. The Faraday cup consists of two parts as it is shown in figures \ref{The Faraday Cup (side view)} and \ref{The Faraday Cup (inside view)}. In Figure \ref{The Faraday Cup (side view)}, the part,  numbered as (1), is the converter and connection part of the Faraday cup, joining the vacuum gradient part. And the second part, numbered as (2), is the head of the Faraday cup. At the most-right of the Faraday cup, there are three BNC connections which are used for different purposes.  The current reading BNC connection of the Faraday cup is  numbered as (3). Both the converter-connection part and the head of the Faraday cup is made of  stainless steel. A copper O-ring is placed between the part (1) and the part (2) while joining them together.

%5.19

The profiles of the parts are shown in Figure \ref{The Faraday Cup (inside view)}. The head of the Faraday cup has an aperture with the diameter of 5 mm. The incoming ion beam impinges on this aperture. The area of this aperture is $0.196 cm^{2}$.

%5.20

In this work, the microwave induced hydrogen plasma was generated at the pressure of 1.5 mbars and the extraction voltage, $V_{ext} $ was applied to the first single aperture at a voltage of $3kV$.  Moreover, the conical single aperture and the frame of the Faraday cup were grounded. The vacuum gradient part was kept at a pressure of order of $10^{-3}$mbars. At this plasma pressure and  extraction voltage, $V_{ext} $, the impinging ion current was measured as  $25 \mu A$, passing through the aperture of the Faraday cup with an area of  $0.196 cm^{2}$. The total current, passing through per centimeter square was calculated as $125 \mu A$.

The trajectory of the generated ions were simulated with the actual dimensions of the  system. The simulations were done with the computer program, known as SIMION Version 8.0.

\begin{figure}[H]
 \centering
 \includegraphics[width=1\textwidth]{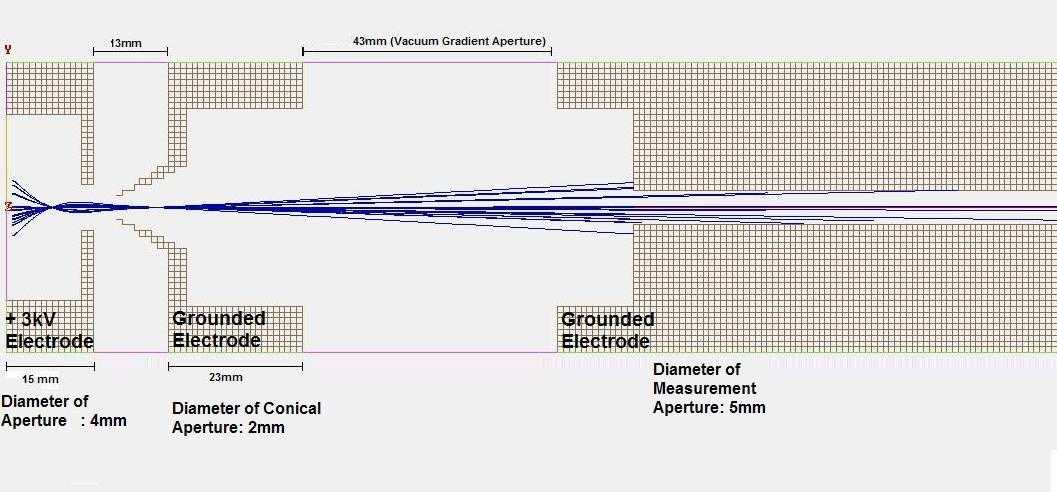}
 \caption{The side view of the simulation of the system}
 \label{The side of the simulation of the system}
\end{figure}

The simulation of the ion trajectories were made for 20 protons, in the vicinity of the first single aperture without any thermal energies. The protons were accelerated by  $3kV$. The aperture of the Faraday cup was located at the most-right end of the simulation screen. For low densities, the focusing of the particles can be achieved by the geometry of the extraction apertures and the extraction electrodes. The side view of the  simulation of the system is shown in Figure \ref{The side of the simulation of the system}.

%5.21

\begin{figure}[H]
 \centering
 \includegraphics[width=0.9\textwidth]{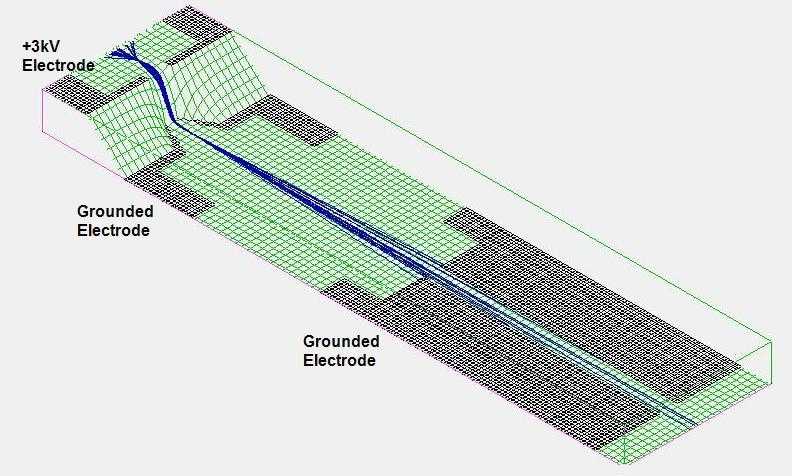}
 \caption{The voltage  topography of the simulated system}
 \label{The voltage  topography of the simulated system}
\end{figure}

 In Figure \ref{The voltage  topography of the simulated system}, the voltage topography of the simulation can be seen; in addition to this, the motion of the particles  can be easily perceived through the voltage downstream.

%5.22

\newpage

% CHAPTER 6
\chapter{DISCUSSION AND CONCLUSION}
\label{chp:conclusion}
In this thesis, the microwave induced plasma ion source was investigated in detail. In order to achieve this,  the ion production via plasma by using a microwave radiation and ion extraction systems were discussed. Moreover, microwave source design and plasma induction systems were studied. In order to obtain and to have the knowledge of the plasma parameters, the double Langmuir probe  techniques were considered and applied. The vacuum gradient was constituted and used in order to increase the mean free path of the extracted ions.

In order to prevent the electrical breakdown  between the extraction electrodes  the estimated Paschen curve for the hydrogen  gas, in Figure 2.4, was drawn and used to verify the clear extraction  without any breakdown. The estimated value of the breakdown voltage is more than 10 kV from the estimated graph  in Figure 2.4, for the minimum separation of 5 mm  and it is also consistent with Equation 6.1 \cite{wolf}.

\begin{equation}
 d\geq 1.41 \times 10^{-2} V
\end{equation}

where $d$ is the separation distance of the extraction electrodes, in  units of millimeter $(mm)$ and $V$ is the extraction voltage, in  units of kilovolt $(kV)$ between the electrodes. For this designed system, the relation was found for the minimum separation gap, $d=5mm$, and the extraction voltage, $V=3kV$, as

\begin{equation*}
 5mm\geq 1.41 \times 10^{-2} (3kV)
\end{equation*}
\begin{equation*}
 5 mm \geq 7.3 \times 10^{-2} mm
\end{equation*}

Our designed and constructed system  worked as it was expected. However, some errors occurred while taking the data at the midpoint of the waveguide. Before discussing the main errors, the known errors should be made clear.

There was an anomaly for the plasma density at the midpoint of the waveguide. The electron density was decreasing, when the probe tips were drawn out of the waveguide as it was expected. However,  the data measurements  at the midpoint  were not accurate. The calculated value of the density is in the order of $ 10^{8}  cm^{-3}$ and it is much  smaller than the values, which were calculated for the dispersion zones. There is a contradiction and, to analyze this situation, a good filtration and an isolation should be done for the double Langmuir probe. When the probe tips were immersed into the waveguide area, the microwave radiation could cause sparks or generate plasma regions in between the pyrex probe tube through which the tungsten probe wires crossing. In addition to these, there could be one more  reason. The density of the particles decreases in a great deal while they spread towards the waveguide borders. And also the TEM mode of the microwave  sustains this distribution while it  decreases its field strength. Moreover, the one of the nodes of the standing wave was close to the pyrex plasma tube,  the tips were  not located accurately. If the tips had been moved a little bit out of its determined location, the data would not been measured accurately.  The dislocating  the tips is the other reason of not measuring the maximum values of the plasma parameters at the midpoint of the waveguide where the plasma parameters reach the maximum values as they are expected to be.    When the probe tips were drawn away from the waveguide borders, where the dispersion began, the electron density decreased as it was expected.

As a result, the  microwave ion source system was designed and constructed by generating the hydrogen plasma. The plasma was generated by using the microwave source with the frequency of 2.45 GHz and the power of 700 Watt. The generated microwave was pulsing with the frequency of 50 Hz. The plasma parameters were taken by the double Langmuir probe and the ion current was measured by the Faraday cup. The designed and constructed   microwave system    generated  the hydrogen  plasma with  $\sim  10^{10}  cm^{-3}$   electron  density   ($n_{e}$)   and   $\sim 2  eV $ electron  temperature  ($T_{e}$)  in  the  pyrex  chamber (tube)  at  the  pressure  of  1.5  mbar  in the region,  out of the microwave-guide, where the plasma dispersed.

%density değeri karşılaştırmasını koy.
By referring Equation  6.2, there is an upper limit for the electron density, $n_{e}$, of the plasma and it is about $n_{e}=6.66 \times 10^{10} cm^{-3}$. For the calculated values from the I-V curves, mentioned above and also in 'Chapter 5', the  maximum value of the electron density was found as  $2\times10^{10} cm^{-3}$ at the outside borders of the waveguide, so the density values verify the theoretical inequality, mentioned below. And the density also decreases rapidly in few millimeters from  $10^{10} cm^{-3}$ to $10^{9} cm^{-3}$ order.
%eq 2.21
\begin{equation}
 n_{e}(=n_{i})\leq 1.11 \times 10^{10} f^{2} 
\end{equation}

\begin{equation*}
 2\times10^{10} cm^{-3}\leq 6.66 \times 10^{10} cm^{-3} 
\end{equation*}

For the   ion extraction part, an ion current was generated by the designed system and the measurements were done by the Faraday cup.
The ion current was measured as 125  $\mu A/cm^{2}$  at  a distance of approximately 125 mm away from the second extraction electrode by  applying  the  extraction  voltage  of  3 kV to the first single aperture electrode and grounding the conical single aperture and the frame of the Faraday cup.

% simulaston ile ilgili olarak
In Figure 6.1 \cite{brown}, three  ion extraction processes are given with different plasma densities. The densities increase as $n_{1} \leq n_{2} \leq n_{3}  $ and the extraction voltage is held constant. If the plasma density is too low then the plasma border (plasma meniscus) forms a concave shape  with the extraction voltage.  With  increasing density this concave shape turns out to  be  a convex shape. After the second electrode, the ions travel a distance, they repel each other and there occurs a divergence because  the electric field is created by  the ions. As a result, the ions repels each other due to this electric field.  

\begin{figure}[H]
 \centering
 \includegraphics[width=0.6\textwidth]{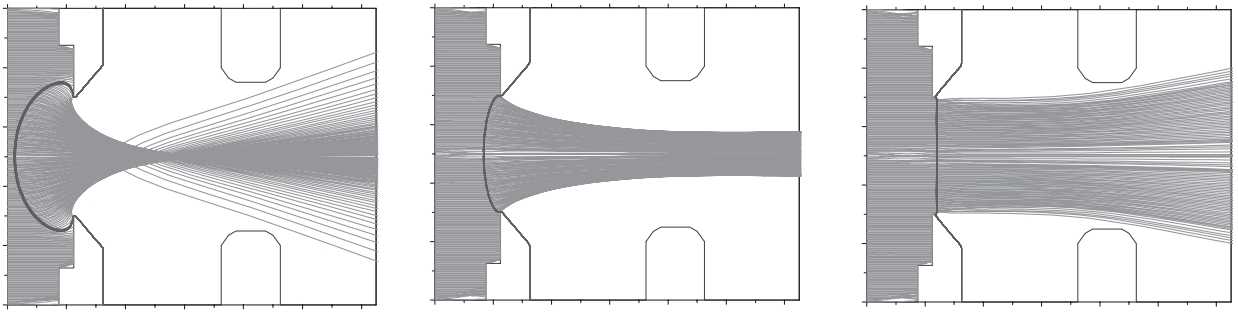}
 \caption{AXCEL-INP simulation for  a diode system with three plasma density and the same voltage drop. From left to right:$ n_{1} \leq n_{2} \leq n_{3}$ [11]}
\end{figure} 

In  figures \ref{The side of the simulation of the system} and \ref{The voltage  topography of the simulated system}, the convex shape is seen clearly before they leave the second electrode and the divergence is formed after the second electrode. This divergence can be controlled by a triode extraction system  and  by an Einzel lens. If a collimation is done by this way, the beam quality can be increased. The Einzel lens system should be added to the beam line to provide the collimation through the  beam line.

% buraya perveance ile ilgli hesap filan koy.
For the designed system, the parameters  of the ions, such as the  perveance, $\mathcal{P}$, so the ion current, $I_{i}$  should be compared with the value of the maximum perveance,  $ \mathcal{P}_{max} $, so the maximum extractable ion current, $I_{imax}$. The values of the parameters are calculated by 

\begin{equation}
 \mathcal{P}_{max}=\frac{ \pi \varepsilon_{0}}{9} \sqrt{ \frac{2(1.6\times10^{-19}C)}{(1.67 \times 10^{-27}kg)}} \left(\frac{1mm}{13mm} \right)^2
\end{equation}
\begin{equation*}
 =2.53\times10^{-10} (A/V^{3/2})
\end{equation*}

If the maximum perveance, $ \mathcal{P}_{max} $, is multiplied by the extraction voltage, $V_{0}$, of 3000 volts, the maximum ion current, in  the vicinity of the conical single aperture, is found: 

\begin{equation}
 I_{imax}= 2.53\times10^{-10} \times 3000^{3/2}
\end{equation}
\begin{equation*}
 \qquad= 4.16\times 10^{-5}(A)=41.6(\mu A)
\end{equation*}

The measured value of the ion current is $25 \mu A$. If the  theoretical and measured values are compared,  the ion efficiency,  $\eta_{a-a}$, between the conical single aperture and the aperture of the Faraday cup is found as:

\begin{equation}
 \eta_{a-a}=\frac{25}{41.6}
\end{equation}
\begin{equation*}
 \quad=0.60
\end{equation*}

The percentage of the ion efficiency is  60\%. Also it can be seen from Figure \ref{The side of the simulation of the system} that 12 of 20 particles can reach  into the middle of the grounded electrode where the measurements were taken. The measured value, above, is consistent with the  value, which is found from Figure \ref{The side of the simulation of the system}. These values, for the ion current readings, were done without applying any Einzel lensing or any collimation process for the extracted ion beam. 

For the further research, the relations between the microwave plasma parameters and the perveance parameter can be investigated by changing the extraction voltage, plasma pressure. In addition to them, the design of the extraction apertures can be changed and effects on the perveance can be investigated further.

\newpage

\bibliographystyle{unsrt}
\bibliography{thesis.bib}
%\input{references.tex}

% if you have more that one appendix, then use \appendices, otherwise use \appendix

%\appendix
% \appendices

%\input{appendix1/appendix1.tex}

% \input{appendix2.tex}

% \input{appendix3.tex}

\end{document}